\documentclass[preprint2]{aastex}

\usepackage{natbib}

\newcommand{\lsim}{\,\rlap{\raise 0.35ex\hbox{$<$}}{\lower 0.7ex\hbox{$\sim$}}\,}
\newcommand{\gsim}{\,\rlap{\raise 0.35ex\hbox{$>$}}{\lower 0.7ex\hbox{$\sim$}}\,}

\def \msun{{\,M_\odot}}

\def \dg{^{\circ}}

\def \8{SN~1987A}
\def \kms{\rm{km\ s^{-1}}}

\shorttitle{Morphology of the ejecta in Supernova 1987A}
\shortauthors{Larsson et al.}

\begin{document}

\title{The morphology of the ejecta in Supernova 1987A: a study over time and wavelength}


\author{Josefin~Larsson\altaffilmark{1}, Claes~Fransson\altaffilmark{2}, Karina~Kjaer\altaffilmark{3},  Anders~Jerkstrand\altaffilmark{4}, Robert~P.~Kirshner\altaffilmark{5},  Bruno~Leibundgut\altaffilmark{3}, Peter~Lundqvist\altaffilmark{2}, Seppo~Mattila\altaffilmark{6}, Richard~McCray\altaffilmark{7}, Jesper~Sollerman\altaffilmark{2}, Jason~Spyromilio\altaffilmark{3} and J.~Craig~Wheeler\altaffilmark{8}}


\altaffiltext{1}{KTH, Department of Physics, and the Oskar Klein Centre, AlbaNova, SE-106 91 Stockholm, Sweden}
\altaffiltext{2}{Department of Astronomy and the Oskar Klein Centre,  Stockholm University, AlbaNova, SE-106 91 Stockholm, Sweden}
\altaffiltext{3}{ESO, Karl-Schwarzschild-Strasse 2, 85748 Garching, Germany}
\altaffiltext{4}{Astrophysics Research Centre, School of Maths and Physics, QueenÕs University Belfast, Belfast BT7 1NN, UK}
\altaffiltext{5}{Harvard-Smithsonian Center for Astrophysics, 60 Garden Street, MS-78, Cambridge, MA 02138, USA}
\altaffiltext{6}{Finnish Centre for Astronomy with ESO (FINCA), University of Turku, V\"ais\"al\"antie 20 FI-21500 Piikki\"o, Finland}
\altaffiltext{7}{JILA, University of Colorado, Boulder, CO 803090440, USA.}
\altaffiltext{8}{Department of Astronomy, University of Texas, Austin, TX 78712-0259, USA}

\begin{abstract}

We present a study of the morphology of the ejecta in Supernova
1987A based on images and spectra from the HST as well as integral field
spectroscopy  from VLT/SINFONI. The HST observations were obtained between 1994 - 2011 and primarily probe
the outer H-rich zones of the ejecta.  The SINFONI observations were obtained in 2005 and 2011 and instead probe the [Si~I]+[Fe~II]   emission from the inner regions. We find a strong
temporal evolution of the morphology in the HST images, from a roughly
elliptical shape before $\sim 5,000$ days, to a more irregular,
edge-brightened morphology with a 'hole' in the middle thereafter.
This transition  is a natural
consequence of the change in the dominant energy source powering the ejecta, from radioactive decay before  $\sim 5,000$ days to X-ray input from the
circumstellar interaction thereafter. The [Si~I]+[Fe~II]   images display a more uniform morphology, which may be due to a remaining significant contribution from radioactivity in the inner ejecta and the higher abundance of these elements in the core.  Both the H$\alpha$ and the [Si~I]+[Fe~II]   line profiles show that the ejecta are distributed fairly close to the plane of the inner circumstellar ring, which is assumed to define the rotational axis of the progenitor star.  The  H$\alpha$ emission extends to higher velocities than [Si~I]+[Fe~II], as expected from theoretical models. There is no clear symmetry axis for all the
emission. Instead,
we find that the emission is concentrated to clumps and that the emission is
distributed somewhat closer to the ring in the north than in the
south.  This north-south asymmetry may be partially explained by dust absorption. We compare our results with explosion models and find some qualitative agreement, but note that the observations show a higher degree of large-scale asymmetry.

\end{abstract}


\section{Introduction}

Both observations and numerical simulations show that supernova (SN) explosions are asymmetric. In particular, simulations show that large-scale instabilities are likely to play a key role in triggering the explosions (e.g.~\citealt{Kifonidis2006,Hammer2010,Mueller2012}). The instabilities are expected to be reflected in the morphology of the ejecta in the homologous phase, which thus provides one of the few direct diagnostics of the explosion mechanism.

Information about the morphology of core-collapse SNe can be obtained in several different ways. Direct imaging of Galactic SN remnants can give
detailed information about both the abundance and the density
distributions. However, there are very few cases unaffected by the interaction with the circumstellar and interstellar media, or a central pulsar wind nebulae. By far, the most important case is Cas A, where
both X-ray (e.g.~\citealt{Hwang2012}) and optical
(e.g.~\citealt{Fesen2006}) observations have provided important information
about instabilities and large-scale mixing in the explosion. In addition, light echoes of Cas A have revealed evidence for asymmetries and shown that the supernova was a Type IIb \citep{Krause2008,Rest2008,Rest2011}.

Constraints on the global geometry of SN ejecta can also be obtained from polarization studies (see \citealt{Wang2008} for a
review) and from the line profiles in the nebular phase (e.g. \citealt{Maeda2008,Taubenberger2009,Milisavljevic2010}). Both types of studies have revealed strong evidence for asymmetries, but the lack of spatial information means that there are no unique solutions for the 3-dimensional ejecta distributions.

SN 1987A offers a unique opportunity to study the spatial
distribution of ejecta from a core collapse SN. Its location in
the Large Magellanic Cloud, only $\sim 50$~kpc away, together with the fact that the
explosion was seen only 25 years ago (on 1987 February 23), make it possible to spatially
resolve the ejecta and study the morphology in the homologous phase, before the inner ejecta crash into the circumstellar medium.

The structure of the ejecta reflects the progenitor star. There is an inner, metal-rich core and an outer, hydrogen-rich envelope. Observations of emission lines from the core in \8 show that this extends out to $2,000-3,500\ \kms$ \citep[e.g.][and this work]{Meikle1993}. The core and the envelope are assumed to be separated by a steep density gradient, followed by a power-law gradient outside $\sim 4,000\ \kms$\citep{Blinnikov2000}. Only the inner ejecta (the metal core and the inner part of the hydrogen envelope) have a high-enough density to be observed directly in the images. However, from the time of the onset of the collision with the circumstellar medium \citep{Lawrence2000} we know that the tenuous hydrogen envelope extends out to $>20,000\ \kms$.

The circumstellar material around \8 is most apparent as three rings,
inclined by $38\dg$-$45\dg$ with respect to the line of sight \citep{Tziamtzis2011}. The
inner, equatorial ring spans $\sim 1.1\arcsec \times 1.6 \arcsec$ while the
two outer rings, located above and below the equatorial plane, are
about three times larger. The rings are thought to have been created
some $20,000$ years before the explosion, possibly as a result of a
binary merger \citep{Morris2007}. By 1995, $\sim 2,900$
days after the explosion, the outermost part of the ejecta
had reached the equatorial ring and the interaction caused the first
so-called hot-spot to appear \citep{Garnavich1997,Lawrence2000}. The ring has since been
observed to brighten exponentially across the entire electromagnetic
spectrum
(e.g.~\citealt{Groningsson2008b,Racusin2009,Zanardo2010,Dwek2010}),
although there are now signs that the flux has started to level off in the optical
\citep{Groningsson2008b,Mattila2010}.

To study the inner ejecta it is crucial to obtain high enough spatial
resolution to separate it from the circumstellar rings. This
can be achieved with the Hubble Space Telescope
(HST) at optical wavelengths and with the use of adaptive optics from
the ground at near-infrared (IR) wavelengths. We have recently used HST
imaging observations, carried out approximately once per year since 1994 (day $\sim 2,800$), to
measure the optical light curve of the inner ejecta
(\citealt{Larsson2011}). This showed that the flux declined as expected
from radioactive decay of ${}^{44}$Ti until $\sim 5,500$ days after
the explosion, after which the flux started to increase. The fact that ${}^{44}$Ti  is the dominant radioactive isotope at these late times is also supported by the observation of the associated hard X-ray emission lines  \citep{Grebenev2012}.  

It is clear, however, that radioactive decay cannot explain the re-brightening of the ejecta.  Instead, we demonstrated that the most likely explanation for this is the energy input resulting from the X-ray emission produced by the
collision between the outer ejecta and the equatorial ring. In this paper we will assume that interpretation is correct. Because of the brightening it is possible to  carry out a detailed study of the morphology of the ejecta even at very late times.

There were early observations of asymmetries in \8 long before the
spatially resolved images of the ejecta were obtained with the HST.
Between days $20-100$, fine-structure appeared in the spectra
\citep[the so-called Bochum-event,][and references
  therein]{Hanuschik1990}, which was interpreted as being due to a
high-velocity clump of $^{56}$Ni \citep{Utrobin1995}. Early
speckle observations \citep{Meikle1987,Nisenson1987} and polarimetry measurements \citep{Jeffery1991} also provided evidence for asymmetries. These early observations only
probe the structure of the outer hydrogen envelope and nearby gas though, which is not obviously
connected to the inner ejecta.

More recently, studies of the morphology of the inner ejecta have been carried out
by \cite{Wang2002}, using HST spectral and imaging data from day
$\sim 4,700$, and by \cite{Kjaer2010}, using integral field
spectroscopy in the near-IR obtained with SINFONI at the VLT
around day $6,800$. While \cite{Wang2002} favor a model in which the
ejecta are prolate, with the long axis pointing out of the equatorial
plane, \citet{Kjaer2010} instead find that the ejecta are elongated closer to
the plane of the ring.  The latter conclusion is also supported by observations of light echoes, which probe the first few hundred days after the explosion from different viewing angles \citep{Sinnott2012}.

In this paper we study the morphology of the ejecta using HST imaging
observations obtained between day 2,770 and 8,328, HST/STIS spectral
observations from days 4,571, 6,355 and 8,378, as well as
ground-based VLT/SINFONI integral field spectroscopy observations from
days 6,816 and 8,714. 

This paper is organized as follows: we describe
the observations and data reduction in section 2, present the analysis
of the images and spectra in section 3, 
discuss our results in section 4 and finally present our conclusions in section 5.

\section{Observations and data reduction}

\subsection{HST Imaging}

HST imaging observations of \8 have been carried out regularly since
$\sim 2,700$ days after the explosion (year 1994) as a part of the
Supernova 1987A INTensive Study (SAINTS).  In order to study the temporal evolution of the morphology we selected six observations,
separated by $\sim 1,000$ days, taken in the F675W/F625W
filters. These images are dominated by strong H$\alpha$-emission and offer the best signal-to-noise ratios. 
For two of the selected epochs (day 5,012 and day 8,328) there are also
high-quality images in five other broad filters which we include in
the analysis. The five filters together cover the entire wavelength
interval between $\sim 2,000$ and $\sim 10,000\ \rm{\AA}$, and thus
allow a detailed study of the spectral energy distribution. To show
which lines contribute to the different filters we plot in
Fig.~\ref{throughput} the filter throughput curves superposed on
ejecta spectra from HST FOS-2 G570H (day 3,604, below $\sim 6,800\  \rm{\AA}$) and
HST STIS G750L (day 4,381, above $\sim 6,800 \ \rm{\AA}$).

We used the
  MULTIDRIZZLE\footnote{http://stsdas.stsci.edu/multidrizzle}
software to remove cosmic rays from the images, apply distortion corrections and
combine individual exposures using the drizzle technique
\citep{Fruchter2002}. For each observation we drizzled the images to
an optimal pixel scale based on the instrument and dither pattern
used. The effective spatial resolution of the resulting images is
between $0.06''$ and $0.09''$. A set of common stars located around the remnant were used to align the images, giving excellent accuracy with the positions of stars in the aligned images always agreeing to within $0.02''$. The details of all the observations are
summarized in Table~\ref{imtable}.

\subsection{HST STIS observations}

The HST STIS observations that we use for this work were obtained on  days 4,381 and 4,571 (year 1999), day
6,355 (year 2004) and day 8,378 (year 2010), using the G750M and
G750L gratings. In the wavelength intervals covered by these gratings the H$\alpha$
line dominates the emission from the ejecta. The spectral resolution is  $\sim50\ \rm{km\ s^{-1}}$ for G750M
and  $\sim450\ \rm{km\ s^{-1}}$ for G750L. The details of the
observations are summarized in Table~\ref{stistable}.

For each slit position there are at least three separate exposures
taken at dithered positions along the slit. We combined these using
the CALSTIS software in order to remove cosmic rays and
hot pixels. This successfully cleaned up the data from the first two
observations, but a number of hot pixels remained in the observations
from the later epochs. To remove these we used the IRAF fixpix task
together with a user-supplied bad pixel mask. We then produced
flux-calibrated two-dimensional spectra from the cleaned data with the
CALSTIS x2d task in the standard way, and finally
extracted one-dimensional spectra by summing over rows in the
two-dimensional spectra. All spectra were corrected for the $287\ \kms$ systematic velocity of \8 \citep{Groningsson2008b}.

\subsection{SINFONI observations}
\label{sinfoniobs}

We use integral field spectroscopy observations of \8 obtained
with SINFONI at the VLT on days 6,816, 6824, 6825, 6839 and 6843 (year
2005) and days 8,714 and 8,717 (year 2011). For this work we consider
only the observations taken in the H-band, which is dominated by the
$1.644\ \mu$m line. In the modeling of the  ${}^{44}$Ti-powered spectrum, \cite{Kjaer2010} and \cite{Jerkstrand2011} showed that this line is a
blend of [Si I] and [Fe II], with [Si I] dominating,  and that both lines originate in the regions close
to the inner Fe core. For the two observations that we consider in
this paper the X-ray input may also add a contribution from the
hydrogen envelope which has not undergone nuclear processing.

The SINFONI K band , discussed in
\citet{Kjaer2010} has too low signal-to-noise to be useful for this
analysis, but we note that the morphology seen in this band (dominated by the He I $2.058\ \mu$m line) is
consistent with that seen in  the H-band. The total H-band exposure times are
4,200~s and 3,600~s for 2005 and 2011, respectively. A detailed analysis of the 2005 data has already been presented by
\citet{Kjaer2010} and we refer the reader to that paper for details
regarding the data reduction. The 2011 data were reduced in the same
way, but with a more recent version of the pipeline (v2.2.3, dated
2010-08-12). 

 There are only variable sources within the small field of view of the SINFONI observations, which makes the flux calibration uncertain. To improve the calibration we used  VLT/NACO imaging observations from day 7,201 and 8,644, which have a much wider field of view containing non-variable stars (Mattila et al., in prep.). We extrapolated the continuum level of the SINFONI spectra to account for the small differences in wavelength coverage between the two instruments, and further corrected for the differences in observing epoch by assuming that the flux from the ring evolves in the same way as the soft X-ray light curve  \citep{Helder2013}. The latter assumption is based on the expectation that the H-band flux is likely to correlate with the optical flux, which is observed to correlate with the soft X-ray light curve \citep{Groningsson2008b}. Since there is no published light curve for the ring in the H-band there is a considerable uncertainty in this assumption.  A conservative estimate is that the SINFONI fluxes derived in this way have uncertainties of  $\pm 20 \%$  for day 6,816 and $\pm 15 \%$ for day 8,717. The primary reason for the larger uncertainty  in the former case is the larger difference between the SINFONI and NACO observing epochs. As for STIS, all the spectra were corrected for the $287\ \kms$ systematic velocity of \8.

\section{Data analysis}

\subsection{Imaging}

\begin{figure*}
\begin{center}
\resizebox{\hsize}{!}{\includegraphics{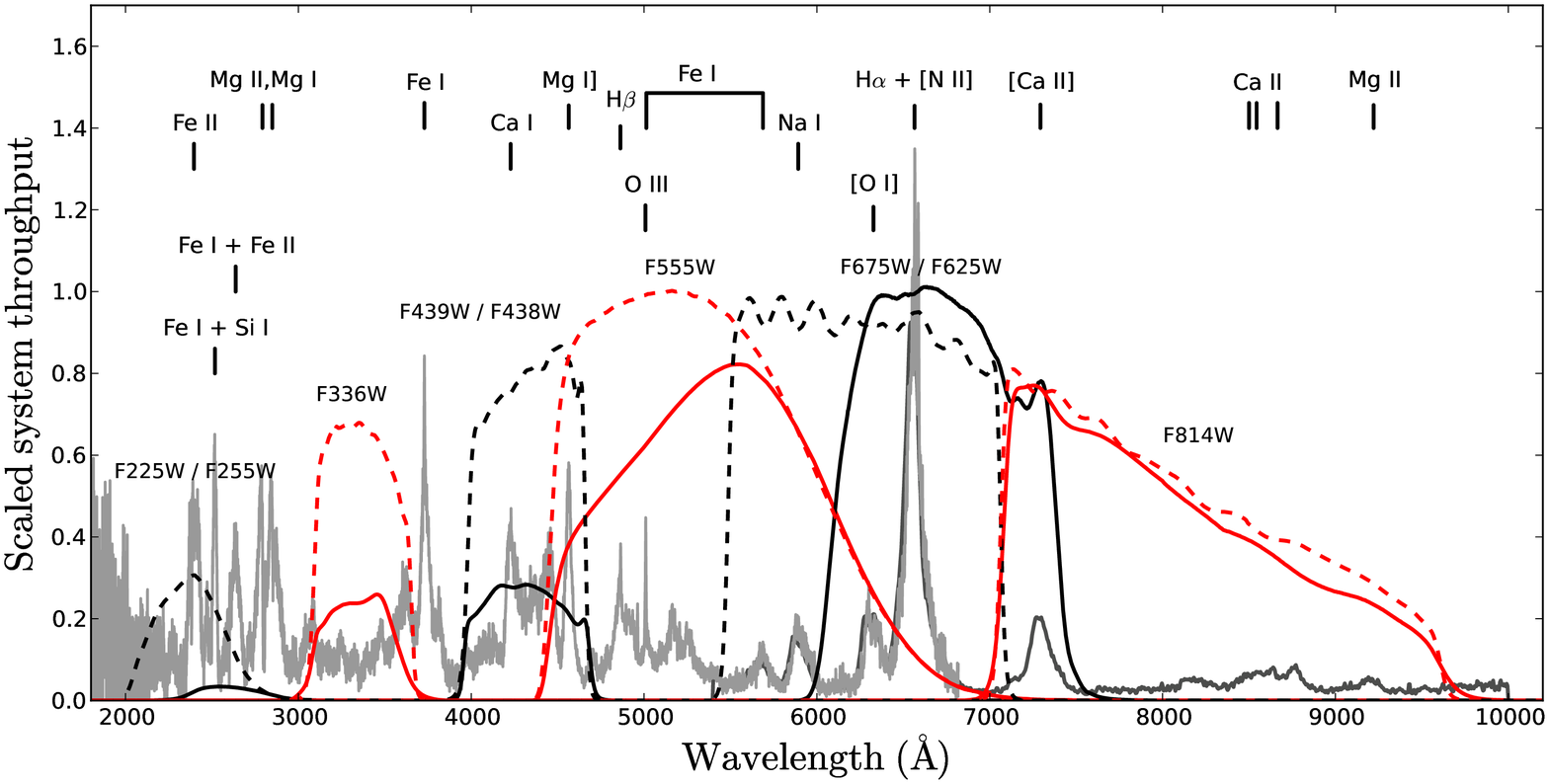}}
\caption{System throughput for the HST filters used in our analysis,
  normalized to the peak value for each instrument. WFPC2 and WFC3
  filters are shown as solid and dashed lines, respectively. Spectra
  of the ejecta from day 3,604 (HST/FOS-2/G570H, light gray) and 4,381
  (HST/STIS/G750L, dark gray) are also shown, with the most important
  lines identified.}
\label{throughput}
\end{center}
\end{figure*}

\begin{figure*}
\begin{center}
\resizebox{76mm}{!}{\includegraphics{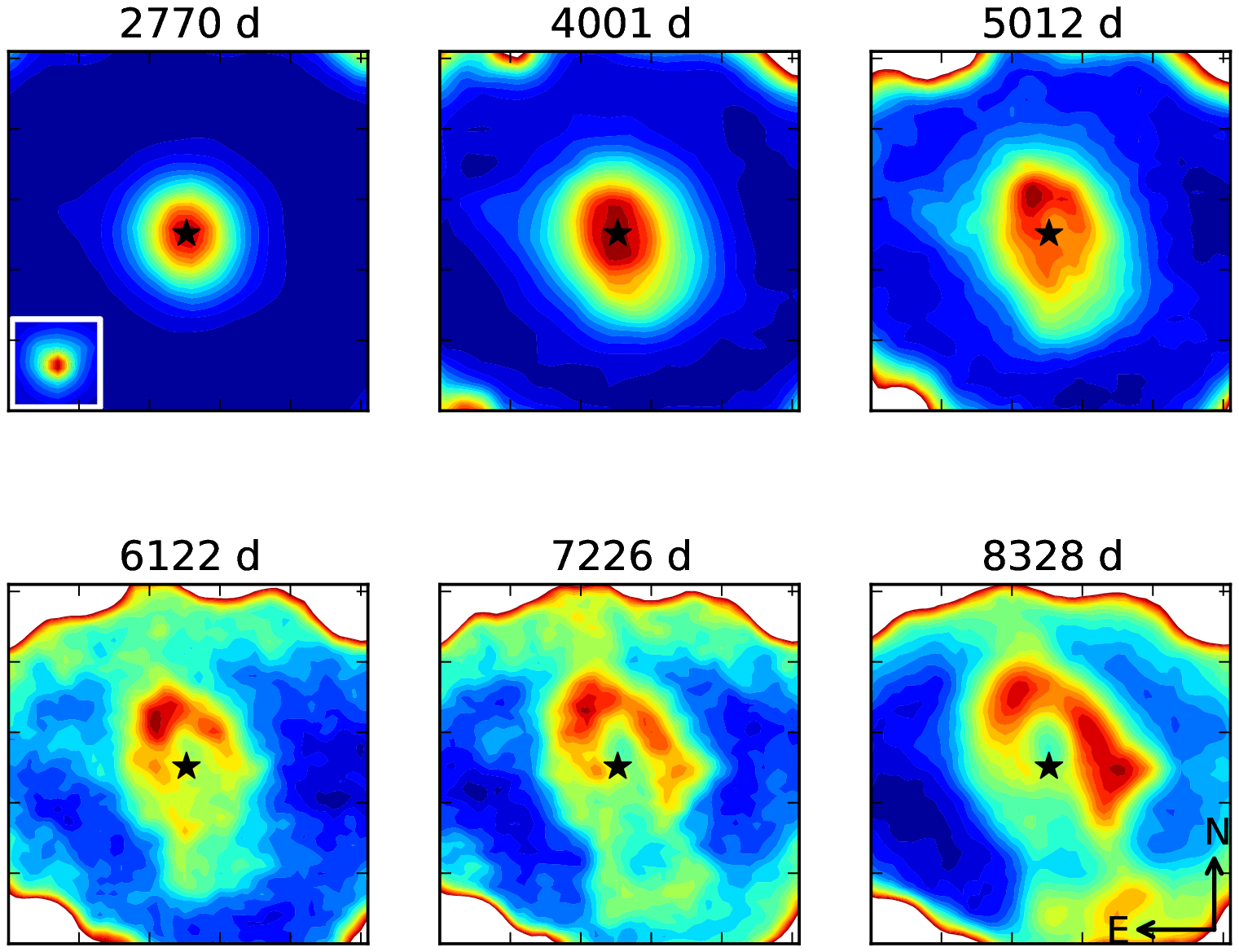}}
\resizebox{76mm}{!}{\includegraphics{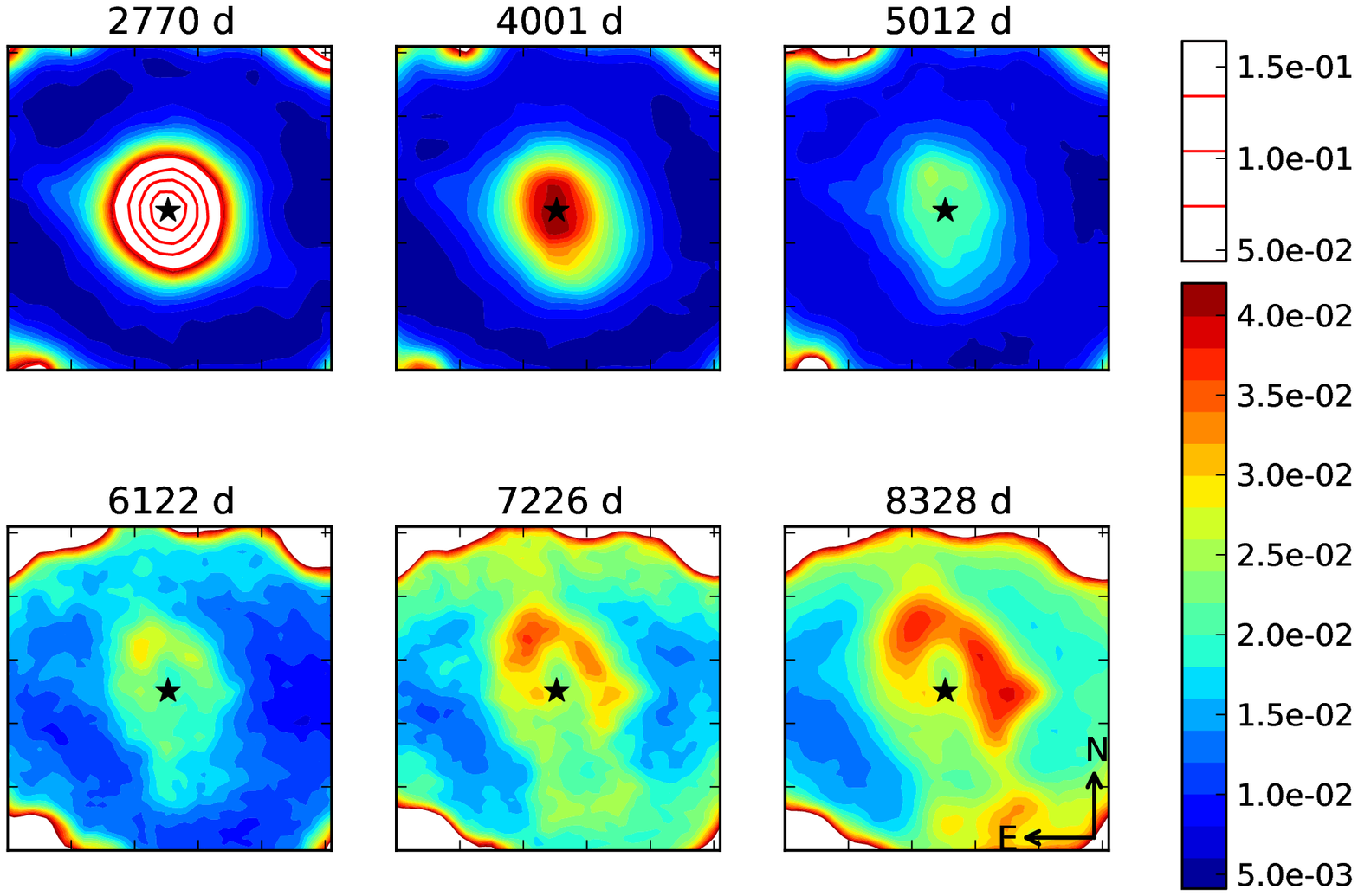}}
\caption{Contour plots of the ejecta produced from HST F675W/F625W
  images. The black star marks the position of the peak brightness in
  the first observation. The contours in the time-series to the left
  are linearly spaced between the minimum (blue) and maximum (red) for each epoch,
  while the images to the right have identical contour levels for all
  epochs.  The red contours on white background correspond to higher
  surface brightness than the filled contours (see intensity scale to the right; the numbers are proportional to the flux per pixel).  To demonstrate that the ejecta are spatially resolved in the first observation we show a nearby star, scaled to the same peak brightness as the ejecta, in the inset in the upper, left panel.  The field of view for each of the images
  is $0.9\arcsec \times 0.9\arcsec$. The first hot spot on the equatorial ring appeared in the
  north-east (outside the region shown here), i.e. in the same quadrant as the ejecta first re-brighten on day 6,122.}
\label{contours}
\end{center}
\end{figure*}

In Fig.~\ref{contours} we show contour plots of the ejecta produced
from the F675W/F625W images in the six selected epochs between day
2,770 and 8,328. In the left panel the contour levels are linearly
spaced between the minimum and maximum surface brightness for each
epoch, while in the right panel the same levels are used for all
epochs. We note that these images are completely dominated by
H$\alpha$-emission (see Fig.~\ref{throughput}). Specifically, we find from the grating spectra that H$\alpha$  makes up more than 70\% of the total broad-band fluxes.  In addition to H$\alpha$ there is a small contribution from Ca~I $\lambda
6572$ (originating in the O/Si and S/Si zones, \citealt{Jerkstrand2011}) and  [O~I] $\lambda \lambda
6300, 6364$. The red wing of the  F675W
filter also has a contribution from [Ca~II]  $\lambda \lambda 7291,7323$, while Na~I $\lambda \lambda 5890, 5896$ influences the blue wing of
the F625W filter. 

 The most important result from Fig.~\ref{contours} is that there is a clear
transition around $5,000-6,000$ days from a roughly elliptical shape
to a more irregular morphology, where the brightest part of the ejecta
is an arc-like structure that is offset from the center of the
remnant. We note that the transition in morphology coincides in time with the
re-brightening of the ejecta. The early images also exhibit an
asymmetry in the north-east and south-west edges of the ejecta due to one
of the outer rings of circumstellar material. The outer rings are
fading with time \citep{Tziamtzis2011} and are not seen at all in the
later images.

\begin{figure*}
\begin{center}
\resizebox{\hsize}{!}{\includegraphics{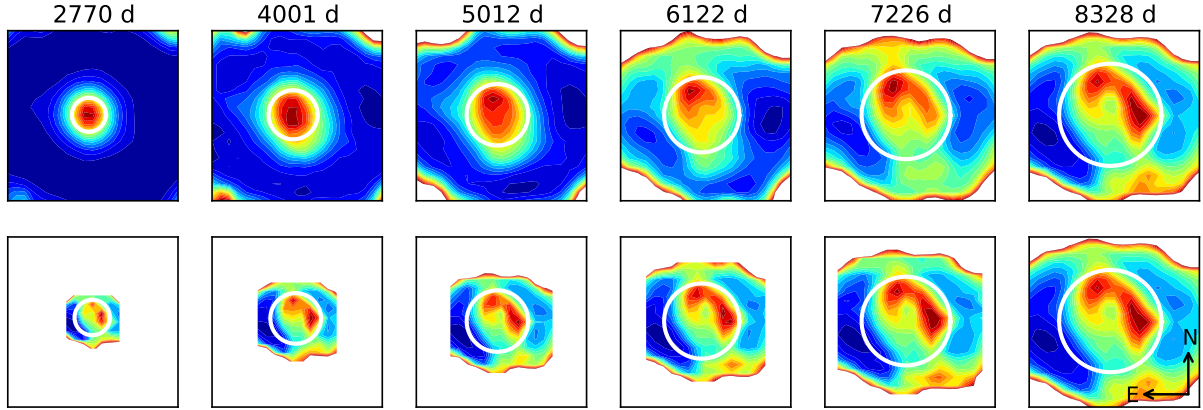}}
\caption{The top row shows contour plots of the ejecta as in
  Fig.~\ref{contours}, but all scaled to the spatial resolution of the
  lowest resolution images. The bottom panel shows the low-resolution version of the image from day
  8,328, reduced in size and "re-observed'' at each epoch, assuming
  that the ejecta were expanding freely.  The contour levels are
  linearly spaced between the minimum (blue) and maximum (red) for each image in
  order to highlight changes in morphology. The white circle is included as a reference point that follows the homologous expansion (with $2,800\ \kms$ in the plane of the sky). The comparison clearly reveals
  that the change in morphology cannot be explained as a result of
  increased resolution due to the expansion. }
\label{shrunk_contours}
\end{center}
\end{figure*}

A natural question to ask is whether the changes in morphology can be
explained simply by the fact that the spatial resolution has increased
with time, primarily due to the expansion of the ejecta but also
because of the change of instruments and dithering strategies. To address
this issue we first degraded the resolution of the later,
higher-resolution images to match the first observation, which has the
lowest resolution. The results are shown in the upper panel of
Fig.~\ref{shrunk_contours}. Next we scaled the final image from day
8,328 to allow a comparison with the previous images. As the ejecta are
expanding homologously, this was achieved by increasing the pixel scale
by a factor given by the ratio between 8,328 days and the number of
days since the explosion for each of the other observations. The results are shown in the bottom
panel of Fig.~\ref{shrunk_contours}. It is clear that the change in
morphology cannot be explained by the increased resolution offered by
the expansion.

\begin{figure*}
\begin{center}
\resizebox{80mm}{!}{\includegraphics{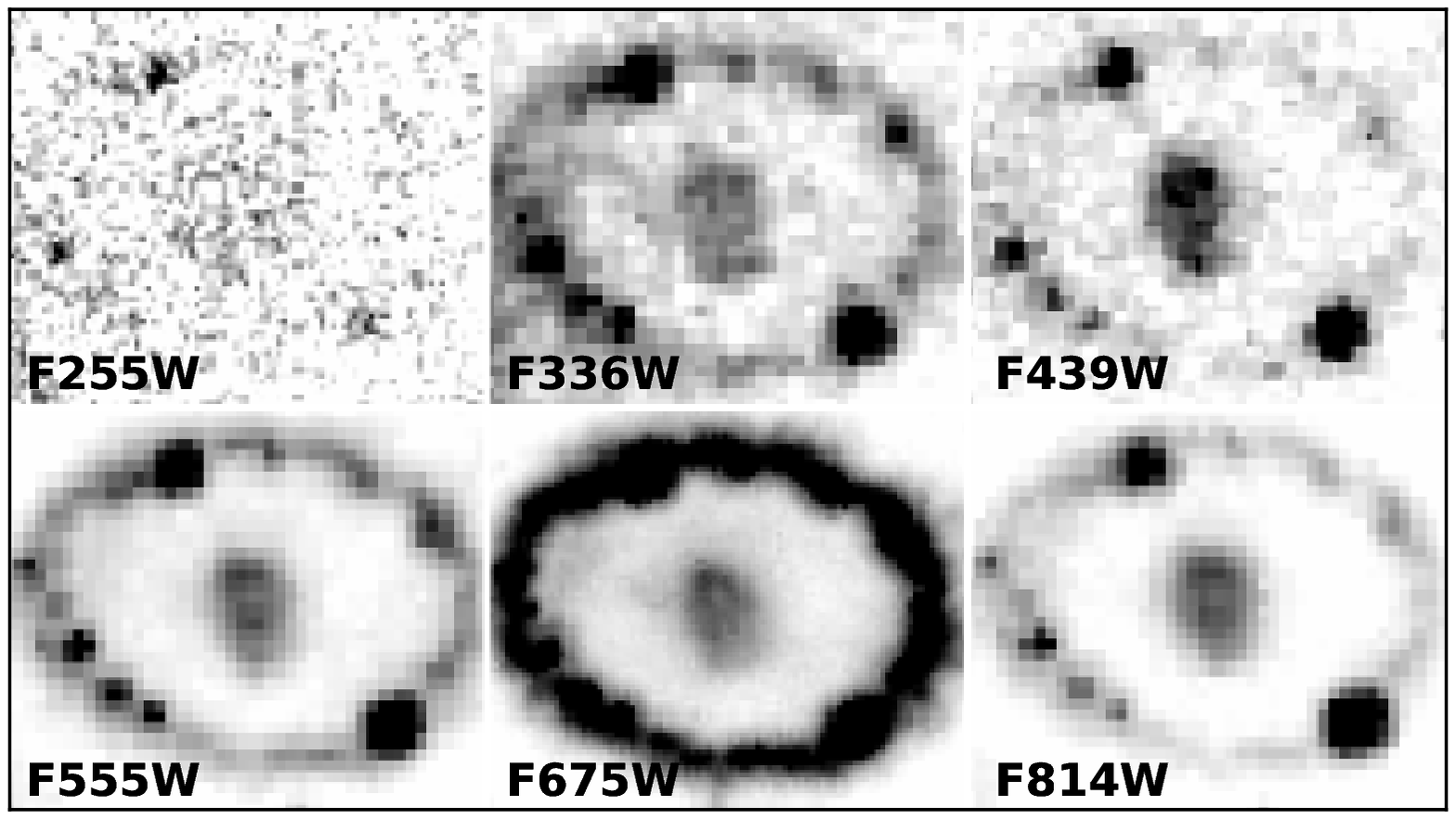}}
\resizebox{80mm}{!}{\includegraphics{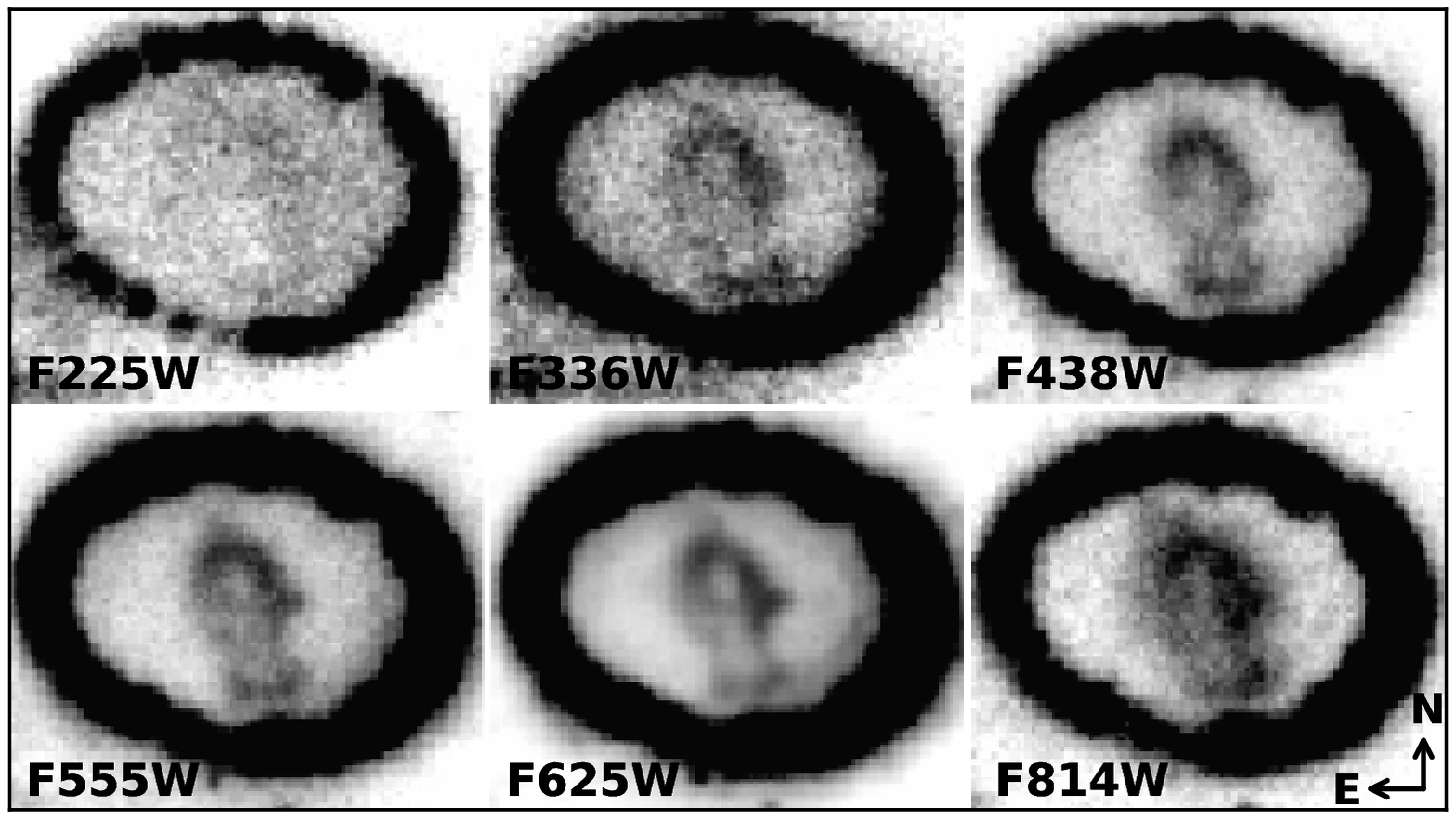}}
\caption{HST images from day 5,012 (left panel, taken with WFPC2) and
  day 8,328 (right panel, taken with WFC3) in six different optical
  filters.  The field of view of each image is $1.9\arcsec \times 1.5\arcsec$. The filter throughput curves are shown in
  Fig.~\ref{throughput}. Note that there are some differences in
  wavelength coverage for especially the F255W/F225W and F675W/F625W
  filters. The minimum and maximum levels in the images were set to
  highlight changes in the morphology. While they change from filter
  to filter, the scales are identical per filter for the two
  epochs. The bright point source in the south-west corner of the
  ring is a star, while the point source in the north-east is the
  first hot spot appearing on the ring. }
\label{filterim}
\end{center}
\end{figure*}

In order to investigate the two morphological stages in more detail we
selected one epoch from each stage (day 5,012 and 8,328, respectively)
for which there is HST imaging observations in six broad filters
covering the entire wavelength interval between $\sim 2,000 -
10,000\ \rm{\AA}$ (Fig. \ref{throughput}).  These images are shown in
Fig.~\ref{filterim}. The images from day 5,012 all display a similar
elliptical morphology, while all the images from day 8,328 reveal a
similar edge-brightened morphology. The only notable exception is that
the ejecta emission in the F225W and F841W images from day 8,328 appear slightly more
extended than in the other images from the same epoch.

To examine whether the spectral energy distribution has changed
between day 5,012 and 8,328 we also measured the flux of the ejecta in all
of the images, the results of which are shown in the left panel of
Fig.~\ref{sed}. The plotted fluxes are the total fluxes in each
filter, i.e. the measured flux density multiplied by the rectangular
bandwidth of the filter and the ratio between the peak throughput and
the throughput at the reference wavelength. We performed the
measurements using elliptical apertures with a position angle of $15
\dg$ and an axis ratio of $0.76$,  approximating the shape of the ejecta. The semi-major axis of the aperture was $0.25''$ for day 5,012
and $0.42''$ for day 8,328, chosen to account for the expansion between
the epochs and to include as much of the ejecta as possible, while
still avoiding the equatorial ring.  This semi-major axis corresponds to $\sim 4,300\ \kms$ in the plane of the sky. The sky background was measured as
an average of five empty regions located around the remnant, and
subtracted from the ejecta fluxes. We note that the sky background is
low, contributing less than $5 \%$ of the flux in any of the
measurements.

A bigger source of uncertainty is the background due to scattered
light from the brightening equatorial ring. While this background is negligible in the early epoch it is fairly high in the late one.  To estimate the contribution from the
ring at 8,328 days we constructed a model for it in the F625W band (where it is
brightest) using the Tiny
Tim\footnote{http://www.stsci.edu/hst/observatory/focus/TinyTim} HST
point-spread-function modeling tool, as explained in
\citet{Larsson2011}.  Using the model we found that the scattered
light from the ring contained within the ejecta aperture corresponds
to $1.3\ \%$ of the flux measured in a $\sim 0.35''$ wide elliptical
annulus centered on the ring itself. Subtraction of this scattered light 
reduces the total flux in the ejecta aperture by about $27\ \%$ in this band. 
We corrected the ejecta fluxes in the other filters by measuring the ring flux in each filter and assuming that the same fraction of this flux ($1.3\ \%$) always falls
within the ejecta aperture.  This correction reduced the ejecta fluxes
by $11 - 27\ \%$ for the different filters. As discussed in
\citet{Larsson2011}, we estimate that the calculation of the amount of scattered light is
accurate to within $15\ \%$ (i.e. the correction in the F625W band is $27\pm4\ \%$ etc.).

The right-hand panel of Fig.~\ref{sed} shows the flux ratios between
the two epochs, divided by the ratios expected due to the small
differences between the filters used for the two epochs
(cf.~Fig.~\ref{throughput}). The latter ratios were calculated using
the SYNPHOT\footnote{http://www.stsci.edu/institute/software\_hardware/stsdas/synphot} 
package with the spectra in Fig.~\ref{throughput} as input. The plot shows that the flux has
increased by factors of $2-3$ in most filters, with the exception of the F255W/F225W band where the increase
is a factor $\sim 5$.

As a final point we compare the HST imaging results with images of the
[Si~I]+[Fe~II]   $1.644~\mu$m line from the SINFONI observations on day
6,816 and 8,714. Contour plots of the SINFONI data covering the $\pm
3,000\ \rm{km\ s^{-1}}$ interval around the line are shown in
Fig.~\ref{sinfcontours}. The plots also show the 50\% and 80\% encircled energy from a point source (calculated as described in \citealt{Kjaer2010}), illustrating the enhanced core and broad wings of the point spread function (PSF). There are some clear differences between Fig.~\ref{sinfcontours} and 
the HST F675W/F625W contour plots in Fig.~\ref{contours}. The peak
brightness of the SINFONI images is close to the center of the
remnant, near the 'hole' in the HST images, and there is also a bright region in the southern part of the ejecta, corresponding to another faint region in the HST images.  Furthermore, there is no strong emission coincident with the bright western region of the ejecta seen in the last HST image.

These differences are even more apparent in Fig.~\ref{comp_contours},
where we over-plot contours from the HST F625W images from days 7,226 and 8,328 on
the SINFONI contour plots from Fig.~\ref{sinfcontours}. The images are well aligned (the
equatorial ring has the same position and shape) and the differences in
observing dates only introduce a $\sim 5$ \%
difference in the size of the ejecta. The spatial resolution is slightly better in the HST images, but this is not enough to
e.g. hide the hole in the SINFONI images. In particular, if the hole
was absent from the SINFONI images due to the broader wings of the
PSFs, we would not expect the brightness to peak near the center of the
hole. The SINFONI images are also consistent with the HST/WFC3 F160W images and the H-band images
obtained with VLT/NACO (Mattila et al., in prep.). We thus conclude that there are significant
differences between the morphology of the $1.644~\mu$m emission and
the H$\alpha$ emission.

\subsection{Spectra}
\label{spectra}

In order to obtain information about the three-dimensional
distribution of material we need to complement the imaging
observations with spectra. Because of the homologous expansion of the
ejecta, the measured Doppler shifts are directly proportional to the
distance along the line of sight from the center of the explosion,
making the spectral information very powerful. For this analysis we
consider HST STIS spectra of the H$\alpha$ line taken on days 4,571,
6,355 and 8,378, as well as SINFONI spectra of the [Si~I]+[Fe~II]  
$1.644\ \mu$m line from days 6,816 and 8,714. We note that there is also
a STIS observation from day 4,381, which was obtained with a wide, $0.5
''$ slit. The full ejecta spectrum from
this observation is shown in Fig.~\ref{throughput}, but we do not
consider it in this section since it is consistent with the spectrum
from day 4,571, which has much better spatial and spectral resolution.

The STIS observations from day 4,571 probe the radioactively dominated
phase. The observations were obtained with the G750M grating for three
adjacent slit positions (slit width $0.1''$) tilted by about $30\dg$
from the north, thus covering most of the ejecta (see
Fig.~\ref{regions99m}). We extracted one-dimensional spectra in four
$0.15''$ regions along each slit as marked in
Fig.~\ref{regions99m}. The resulting spatially resolved profiles are
shown in Fig.~\ref{spres_profiles99m} and the total profile is shown
in Fig.~\ref{full_profiles_stis}. Due to the excellent spectral
resolution we can clearly separate the narrow lines from the outer
ring ([N II] $\lambda6548$, H$\alpha$ and [N II]
$\lambda6583$) from the broad ejecta profile. The latter
extends out to at least $\pm 4,000\ \rm{km\ s^{-1}}$. The spatially
resolved profiles show a trend of more blueshifted emission in the
north and more redshifted emission in the south. The emission is also
slightly brighter in the north. This is reflected in the full profile
in Fig.~\ref{full_profiles_stis} (black line), which is asymmetric with a somewhat
stronger blue wing. This asymmetry may partly be due to the
presence of dust (see further discussion in section \ref{dust}).

Compared to this early observation the STIS observations from days
6,355 and 8,378 have poorer spectral and spatial resolution, as they
were taken with the G750L grating and a slit width of $0.2''$. For day
6,355 three adjacent slit positions oriented in the north-south
direction cover most of the ejecta, while on day 8,378 there is only
one slit position covering the central part (see
Fig.~\ref{regions0410}). As above, we extracted one-dimensional
spectra in $0.15''$ steps along each of the slits
(Fig.~\ref{regions0410}). The results for the two epochs are shown in
Figs.  \ref{spres_profiles05} and \ref{spres_profiles11}, respectively. The 2D STIS
spectra are plotted in Fransson et al. (2013, in prep. F13 from here on), showing both the H$\alpha$ and the [Ca II] emission. 

The SINFONI observations were carried out 461 and 629 days later than
the two STIS observations, which is close enough to make useful
comparisons (the offsets correspond to $\sim 7 \%$ of the age of the
supernova). We therefore include the SINFONI $1.644\ \mu$m line
profiles from days 6,816 and 8,714 in Fig.\ref{spres_profiles05} and
\ref{spres_profiles11}, respectively. In order to extract SINFONI
spectra from the same regions as the STIS spectra we used the
geometrical center of the equatorial ring as a reference point,
noting that the size and shape of the ring is very similar in the
optical and near-IR. The extraction regions are shown superposed on
the SINFONI images of the $1.644\ \mu$m line ($\pm
3,000\ \rm{km\ s^{-1}}$) in Fig.~\ref{regions0410}. There are some
uncertainties in the comparison with the STIS spectra owing to the
different PSFs (the SINFONI PSF has an enhanced
core, with broader faint wings), the positioning of the STIS slits,
the image quality etc; however, these effects are small compared to
the systematic effects introduced by contamination from the
equatorial ring.

At these late epochs, the fading emission from the outer ring, which
was identified in the STIS spectrum from day 4,571, will only make a
small contribution to the ejecta spectra (see light curves in
\citealt{Tziamtzis2011}). On the other hand, the equatorial ring is
much brighter at these times (e.g., Fig.~\ref{filterim}) and it is
likely to contribute to the spectra in Figs.~\ref{spres_profiles05}
and \ref{spres_profiles11}. To get an estimate of the level
of contamination we extracted spectra from the equatorial ring for both SINFONI
and STIS and re-scaled them to the ejecta spectra, as shown in
Fig.~\ref{ringcont}. For SINFONI it is clear that the narrow ($\pm
250\ \rm{km\ s^{-1}}$) central component originates from the ring, but
that there are no other strong lines from the ring that could
contribute to the broad ejecta profile. In the case of the STIS
observations, it is harder to determine the contribution from the ring
due to the lower spectral resolution. The ring spectrum contains
intermediate-velocity lines of [N II] $\lambda 6548$,
H$\alpha$ and [N II] $\lambda 6583$ (e.g.
\citealt{Groningsson2008}), which together make a broad, skewed
component that is likely to contaminate the ejecta spectra between
about $\sim -1000 - +1800\ \rm{km\ s^{-1}}$. There are no strong
emission lines from the ring that could contaminate the broad wings of
the ejecta H$\alpha$-profile outside this interval though.

Inspection of the spatially resolved ejecta profiles in Figs.
\ref{spres_profiles05} and \ref{spres_profiles11} show the same trend
of more blueshifted emission in the north and more redshifted emission
in the south, as already noted for the earlier STIS observations in
Fig.~\ref{spres_profiles99m}. The H$\alpha$ and [Si~I]+[Fe~II]  
profiles mainly differ in their blueshifted emission, with the
H$\alpha$-lines exhibiting smoother wings extending to higher
velocities. These properties will be discussed in more detail in
section \ref{discussion} below.

As the slit positions for the STIS observations on day 6,355 cover most
of the ejecta we also show the full profile from these observations in
Fig.~\ref{full_profiles_stis}, together with the total profile from
day 4,571. Even though the spectra were extracted such that most of the
ejecta would be included for both epochs, it is clear that some
uncertainties are introduced by differences in gratings, slit width
and orientation, as well as the expansion of the ejecta. Nevertheless,
it is interesting to note that the two STIS profiles are very similar despite
the change in morphology that has occurred during the five years
between the two observations. The fluxes are also rather similar,
which is likely due to a combination of the uncertainties mentioned
above, together with the fact that the minimum flux was reached in the
time interval between the observations

The SINFONI data also cover all of the ejecta and we therefore next
consider these observations in more detail. Compared to the STIS
observations, the SINFONI data have the advantage of higher spatial
and spectral resolution. In Fig.~\ref{sinf3d} we plot the
three-dimensional emissivity in the [Si~I]+[Fe~II]   line, obtained by converting the velocity
along the line of sight to a distance from the center for each
pixel, assuming homologous expansion and a distance of 50~kpc to \8. The position of the equatorial ring and our line of sight are
also included as reference points. Animations showing the full 3D view of the emissivity, created with Mayavi \citep{Ramachandran2011}, are included in the online version of the journal.

These plots and animations show that the
emission is concentrated in two ''clumps'', one to the north on the
near side and one to the south on the far side with respect to the plane of the sky. In order to show the large-scale structure as clearly as possible we have chosen to only plot ejecta emission brighter than 3 times the continuum level with velocities lower than $3,500\ \kms$ (excluding the central $\pm 250\ \kms$ along the line of sight). The first condition ensures that the plotted data points are well above the noise level and highlights the location of the brightest emission. We note that choosing a lower cutoff simply makes the observed structures slightly larger. The cuts in velocity were chosen in order to avoid emission from the ring blending with the ejecta emission. The downside of this is that some bright emission in the southern tip of the ejecta is excluded (cf. Fig.~\ref{sinfcontours}) and that there is a small gap in the central region.  We stress that  none of these different cuts change the main conclusion that the emission is concentrated to two clumps. It is also clear that there
is very little change in morphology between the two observations.

From these three-dimensional maps we also calculate the radial
intensity distribution by summing over spherical shells of width
$250\ \kms$. The results are shown in Fig.~\ref{radial_dist_sinf}.
For both epochs the intensity peaks around $2,300\ \kms$ and the width
of the distribution at half maximum is $\sim 1,700\ \kms$. We note
that there are systematic uncertainties in these distributions due to
difficulties in defining the continuum level ($< 15$\ \% uncertainty)
and the fact that the resolution is higher along the line of sight
than perpendicular to it by a factor of $7-9$ (the velocity resolution
perpendicular to the line of sight improves with time due to the
expansion).

The change in the ejecta flux between the two SINFONI observations is a factor of $1.2 \pm 0.3$. While this is consistent with the HST observations, the rather large uncertainties in the flux calibration (discussed in section \ref{sinfoniobs}) means that we cannot exclude the possibility that a significant fraction of the emission at these wavelengths is still due to radioactivity.

\section{Discussion}
\label{discussion}

\subsection{Thermalization of the radioactive and X-ray input}
\label{discussion_thermal}

Our observations provide us with intensity maps in different spectral
lines. Because our main goal is to infer information about the spatial
distribution of the different elements, we need to
discuss the connection between the energy input in the form of radioactivity
and X-rays, the density distribution, and the observed
emission. This is discussed in detail in e.g.,
\cite{Kozma1992}, \cite{Jerkstrand2011} and F13, so here we only summarize the main
points.

The decay of ${}^{44}$Ti produces gamma-rays and positrons. \cite{Jerkstrand2011} estimate an optical depth to the gamma-rays
 of $\tau_\gamma \approx 0.05 (t / 8 \ {\rm yrs})^{-2}$. At 20 years $\tau_\gamma $ is only $0.01$, which mens that the positrons dominate the energy input. With full
 trapping of the positrons, as indicated in \cite{Jerkstrand2011},
 almost all radioactive input will be deposited in the
 Fe-rich material. The energy of the
 positrons is deposited as ionizations, excitations and heating,
 resulting in optical and IR emission lines. Because the energy input
 mainly occurs in the Fe-rich material, with a minor contribution from
 the Si/S zone, most of the resulting emission will come from ions of the Fe-rich
 material.

In addition to the energy input directly connected to the ${}^{44}$Ti
decay, there is an important contribution from freeze-out emission
in the hydrogen envelope, where the low density makes recombination
slow and energy input from earlier radioactive decays of ${}^{56}$Co
and ${}^{57}$Co dominate \citep{Fransson1993}. This component mainly results in Ly$\alpha$, two-photon and Balmer line emission.

Even in the H and He zones the optical depth to the UV resonance lines from Mg I-II, Ti I-II, Fe
I-II and other trace elements is  large enough to scatter nearly all of the radiation
emitted in the UV. Some of this will escape as blended lines at longer
wavelengths in the UV and near-UV, while some is emitted as
fluorescence radiation in the optical and IR. Spatially, this emission
comes from the hydrogen envelope, even though the primary origin may
be from the processed Fe core.

At later epochs,  when the ejecta are powered by the X-ray emission, the thermalization of the fast photoelectrons
created by the X-ray absorption is similar to that of the
gamma-rays and positrons resulting from the radioactive decay.  As long as the energy of the photoelectron
is larger than $\sim 0.3$ keV, the fractions of the energy going into heating,
ionizations and excitations are independent of energy
\citep[e.g.][]{Xu1991}. The fractions are, however, sensitive to the ionization level, which determines the efficiency for Coulomb
heating. Estimates in F13 show that the ionization
in the outer regions close to the ring can be high, $\ga 0.1$, while
that in the core is typically $\sim 10^{-3}$. For
ionization fractions in this range, the total fraction of the X-ray energy
emerging as H$\alpha$ is  $2-5 \%$ \citep[][]{Xu1991,Kozma1992}.
Most of the remaining energy emerges as  two-photon emission and Ly$\alpha$. 
In the same way as described above, this
emission may be scattered by UV resonance lines.

\subsection{Evolution of the spectral energy distribution}
\label{discussion_sed}

\begin{figure*}
\begin{center}
\resizebox{80mm}{!}{\includegraphics{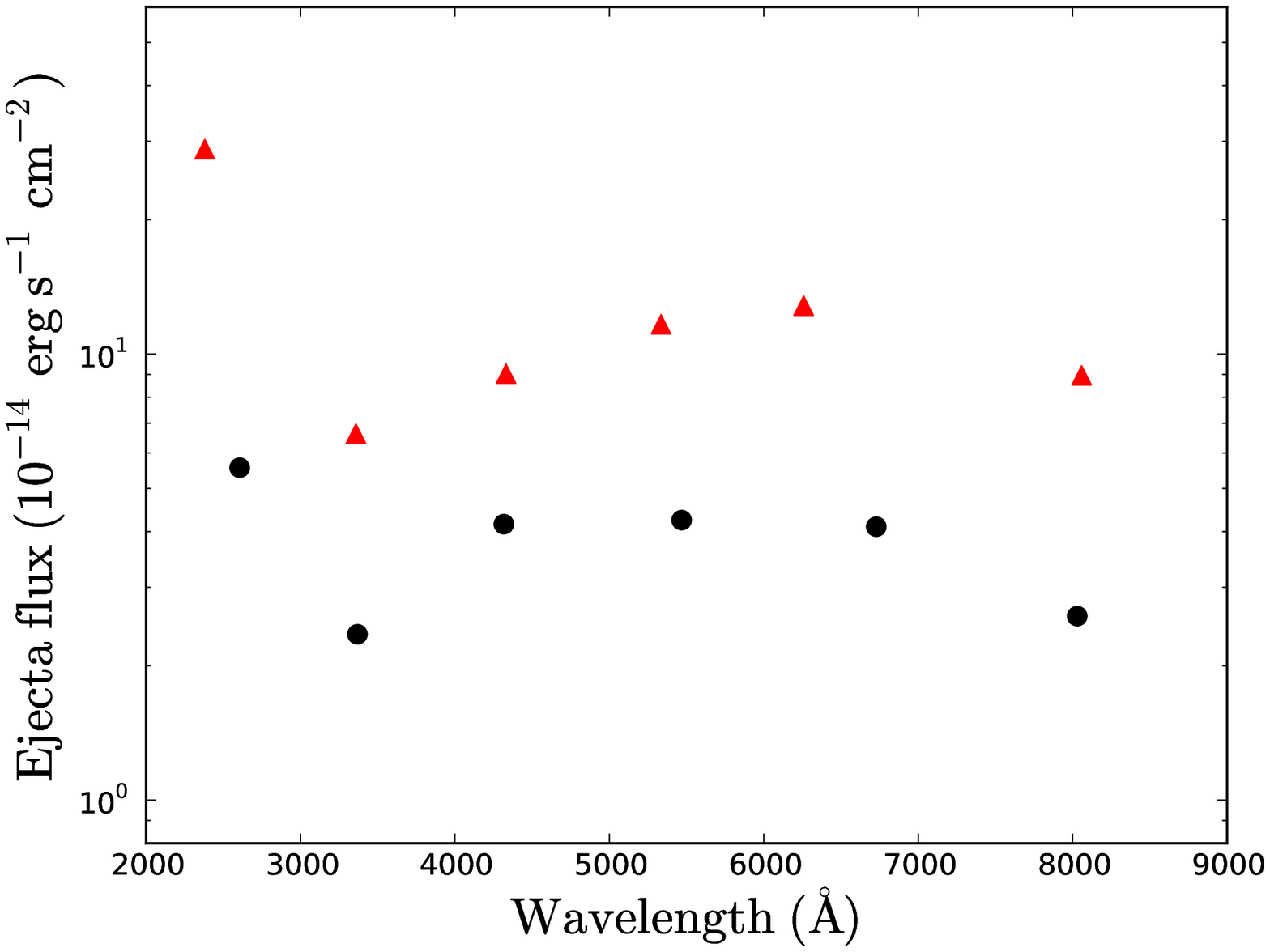}}
\resizebox{80mm}{!}{\includegraphics{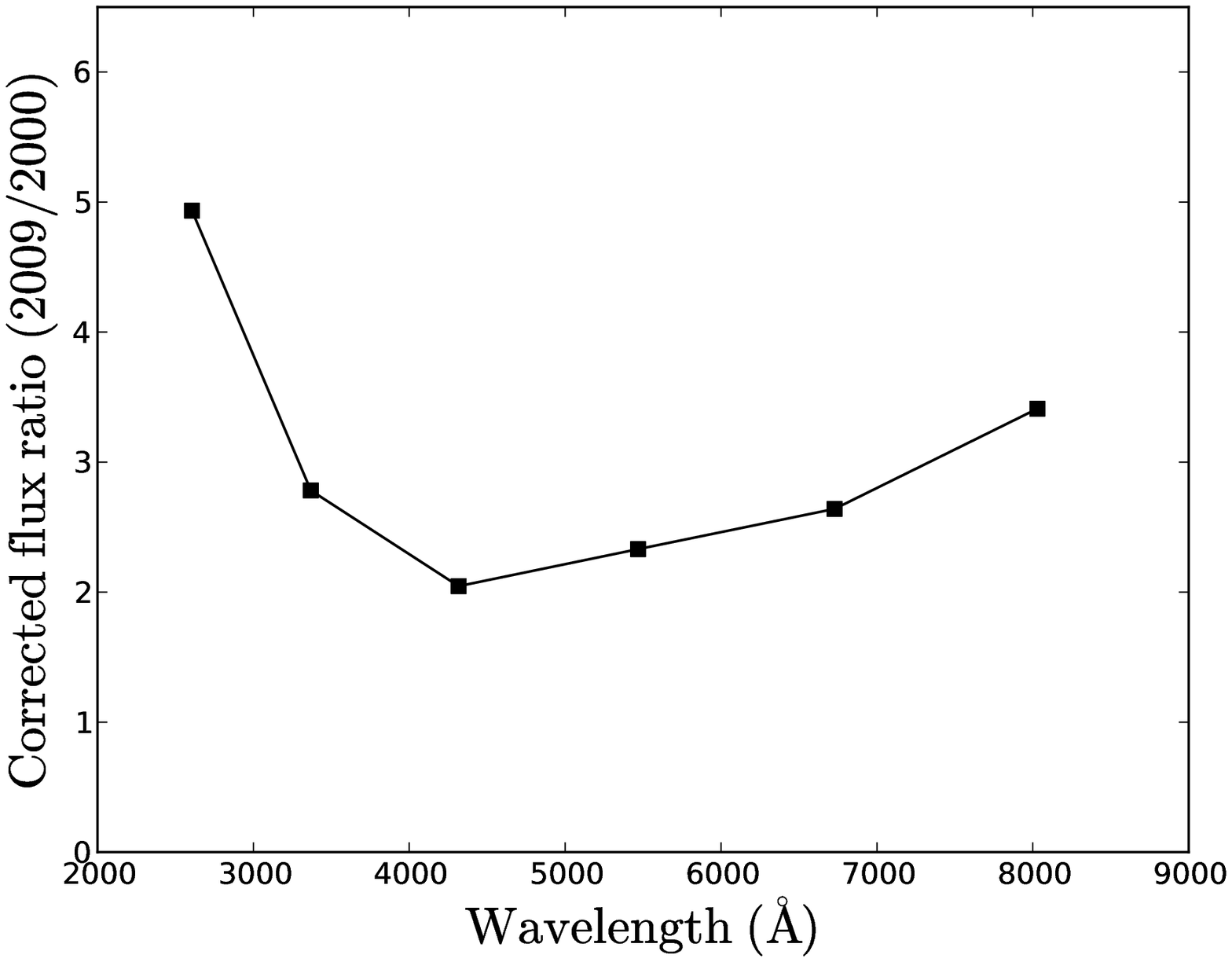}}
\caption{The left panel shows ejecta fluxes on day 5,012 (black circles) and
  day 8,328 (red triangles) for all the images plotted in
  Fig.~\ref{filterim}. The fluxes from day 8,328 have been corrected
  for scattered light from the ring. The statistical errors are about the same size as the plot symbols.  The fluxes have been de-reddened by $\rm{E}_{B-V} = 0.19$ mag using the \cite{Cardelli1989} reddening law (see \cite{France2011} for a motivation of this). The right panel shows the ratio of the fluxes, corrected
  for the fact that the filters used for the two epochs have slightly
  different wavelength coverage and throughputs
  (cf.~Fig.~\ref{throughput}). }
\label{sed}
\end{center}
\end{figure*}

\begin{figure*}
\begin{center}
\resizebox{100mm}{!}{\includegraphics{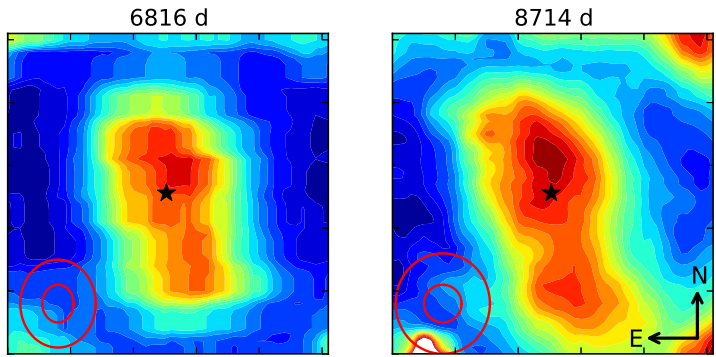}}
\caption{Contour plots of the ejecta produced from SINFONI images of
  the [Si~I]+[Fe~II]   $1.644\ \mu$m line in the $\pm 3,000
  \rm{km\ s^{-1}}$ range. The contours are linearly spaced between the
  minimum (blue) and maximum (red) for each epoch. The black star marks the
  position of the peak brightness of the first HST image from day 2,770
  (cf.~Fig.~\ref{contours}).  The red ellipses in the bottom, left corners show the 50\% and 80\% encircled energy area from a point source. The field of view is $
  0.9\arcsec \times 0.9\arcsec$.}
\label{sinfcontours}
\end{center}
\end{figure*}

\begin{figure*}
\begin{center}
\resizebox{100mm}{!}{\includegraphics{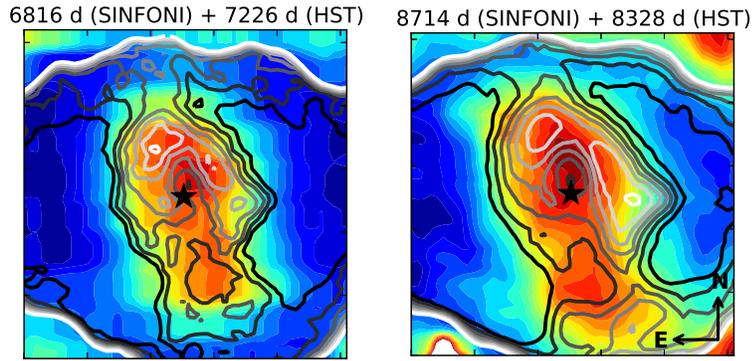}}
\caption{Comparison of contour plots produced from SINFONI [Si~I]+[Fe~II] images
  (in color)  and HST F625W images (grayscale). The SINFONI contours are the same as in  Fig.~\ref{sinfcontours}, while the HST contours correspond to
  every other level above the faintest 25\% in the left
  time-series in Fig.~\ref{contours}. The field of view is $
  0.9\arcsec \times 0.9\arcsec$.  Note that the faintest, central parts of the HST images are filled in by bright emission in the SINFONI images. }
\label{comp_contours}
\end{center}
\end{figure*}

\begin{figure*}
\begin{center}
\resizebox{75mm}{!}{\includegraphics{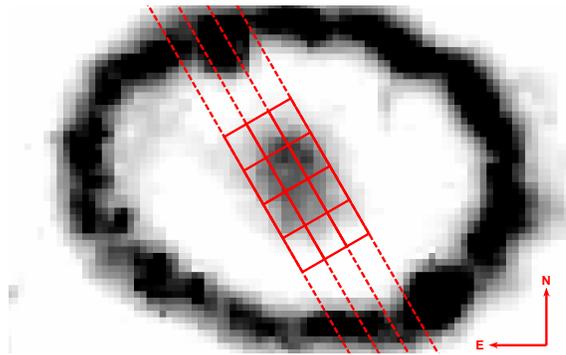}}
\caption{Slit positions used for the STIS observations on
  day 4,571, shown as dashed lines superposed on the WFPC2 F675W
  image from day 4,727. The slit width is $0.1 ''$ and the field of view is $2.2\arcsec \times 1.4\arcsec$. The extraction
  regions used for producing the spectra in
  Fig.~\ref{spres_profiles99m} are shown as solid rectangles (size $0.1\arcsec \times 0.15\arcsec$). The $0.5
  ''$ slit that was used for the STIS observations on day 4,381
  (spectrum shown in Fig.~\ref{throughput}) has nearly the same
  orientation as the $0.1 ''$ slits plotted here.}
\label{regions99m}
\end{center}
\end{figure*}

\begin{figure*}
\begin{center}
\resizebox{\hsize}{!}{\includegraphics{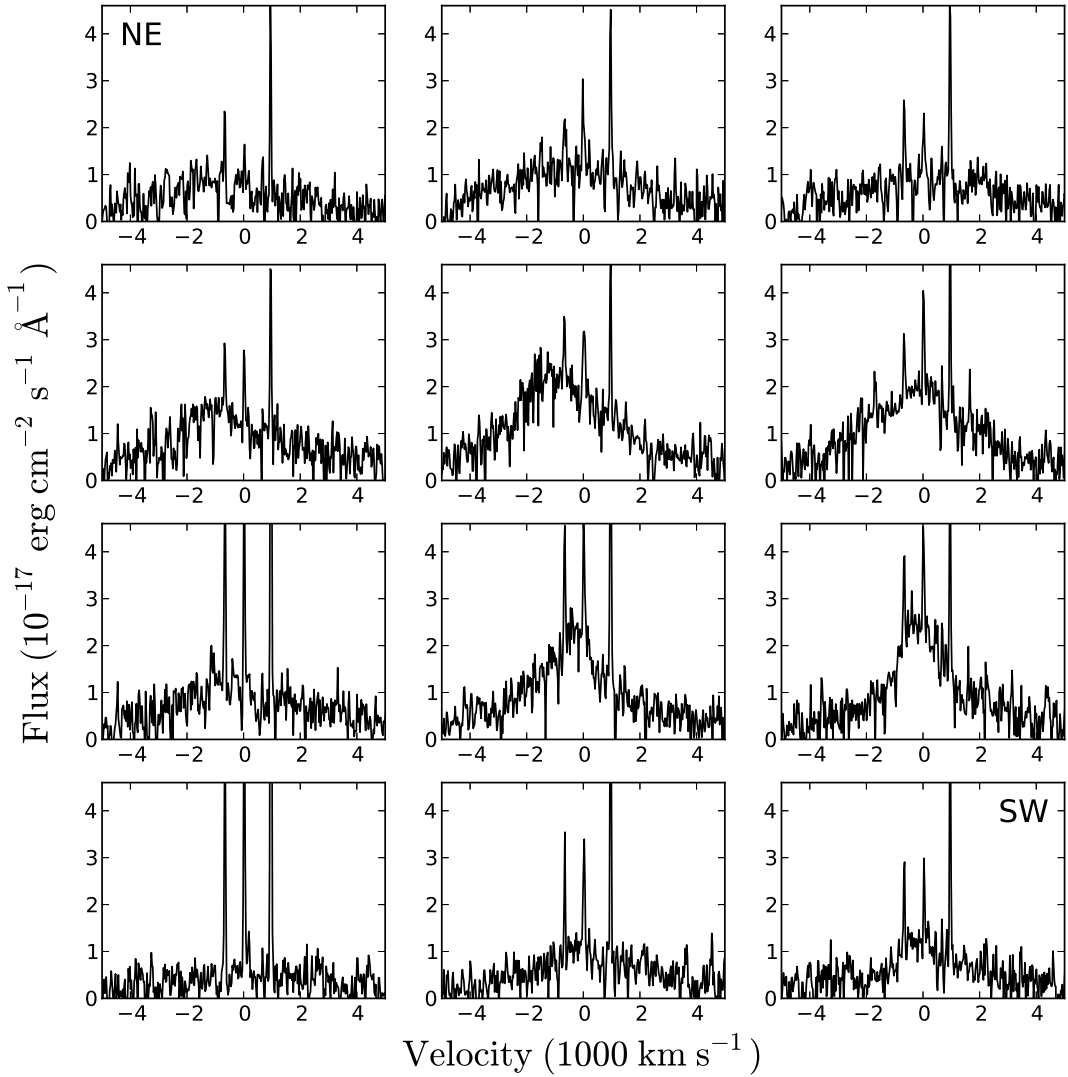}}
\caption{Spatially resolved STIS G750M spectra from day 4,571. The spectra
  are centered on zero velocity for H$\alpha$.  Each spectrum was
  extracted from a $0.1\arcsec \times 0.15\arcsec $ region, as shown in
  Fig.~\ref{regions99m}. The narrow lines originate from one of the
  outer rings, which is partly projected on top of the ejecta. }
\label{spres_profiles99m}
\end{center}
\end{figure*}

\begin{figure*}
\begin{center}
\resizebox{80mm}{!}{\includegraphics{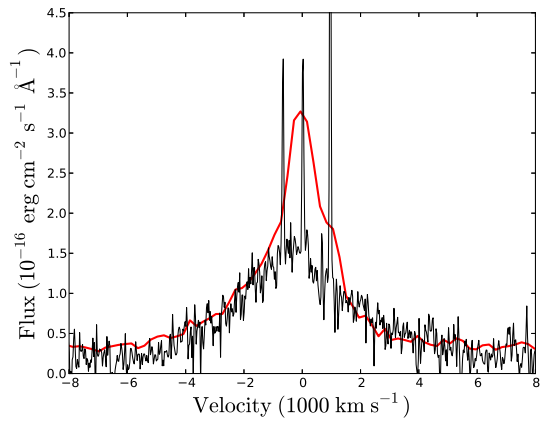}}
\caption{Comparison of ejecta H$\alpha$ profiles from HST STIS
  observations obtained on day 4,571 (G750M grating, black line) and
  day 6,355 (G750L grating, red line). The spectra were extracted from
  the total regions marked with solid lines in Fig.~\ref{regions99m}
  and the top, left panel of Fig.~\ref{regions0410}, respectively. The
  extraction regions cover most of the bright, central ejecta for both
  epochs, although the differences in  resolution (nearly a factor 10 better in the early observation), slit width and orientation, as
  well as the expansion of the ejecta, introduce some uncertainties in
  the comparison of the full profiles. }
\label{full_profiles_stis}
\end{center}
\end{figure*}

\begin{figure*}
\begin{center}
\resizebox{75mm}{!}{\includegraphics{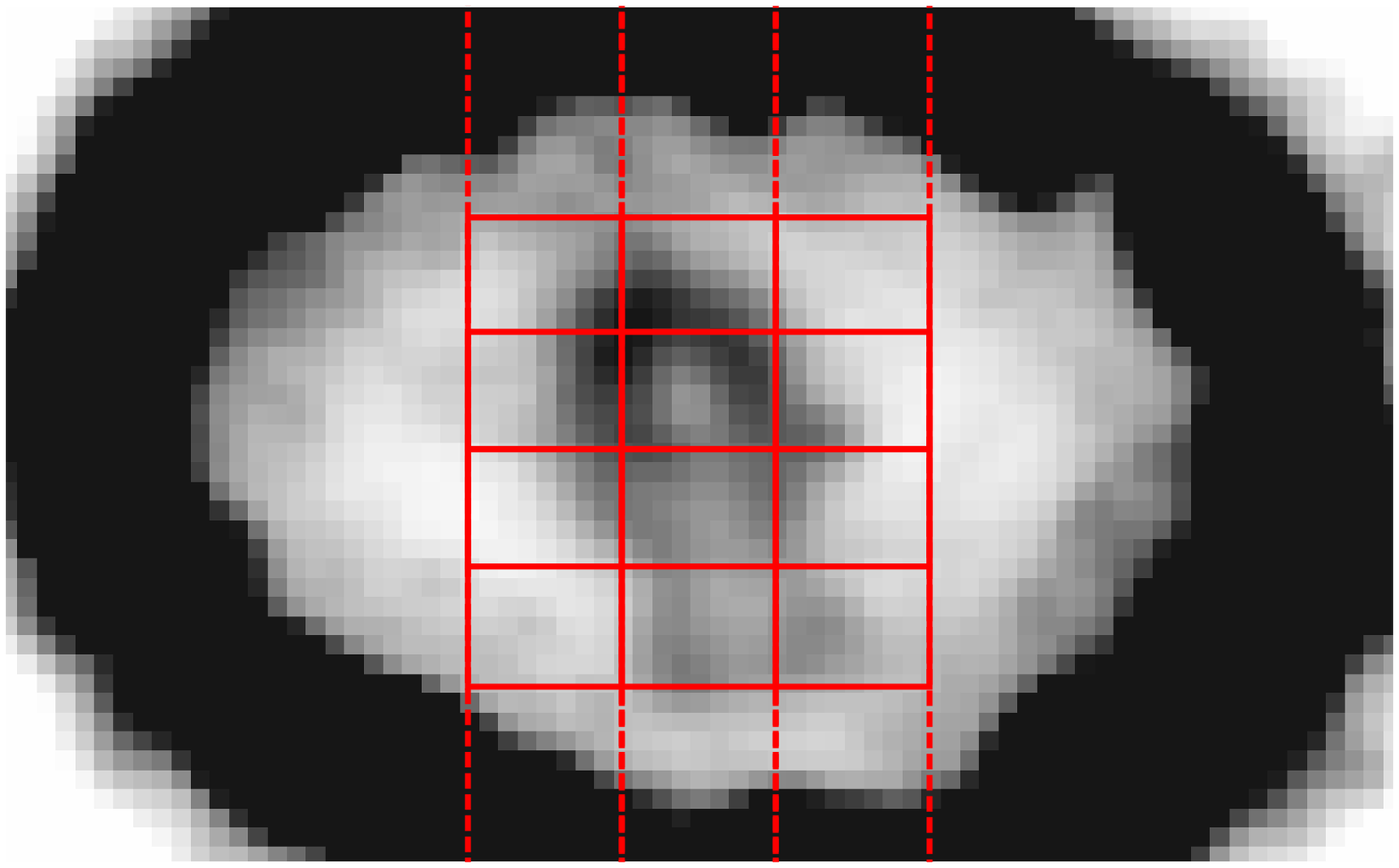}}
\resizebox{75mm}{!}{\includegraphics{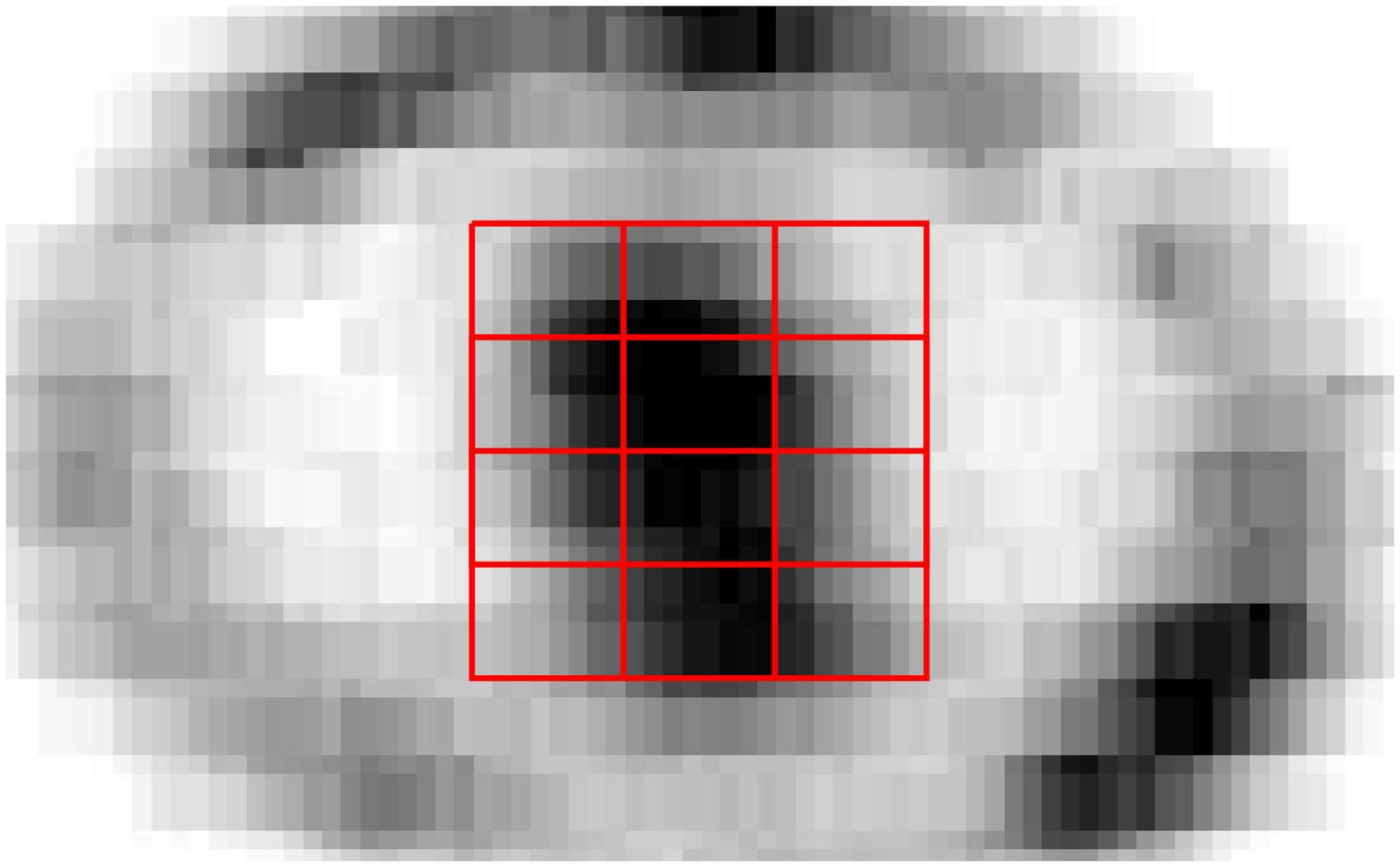}}
\resizebox{75mm}{!}{\includegraphics{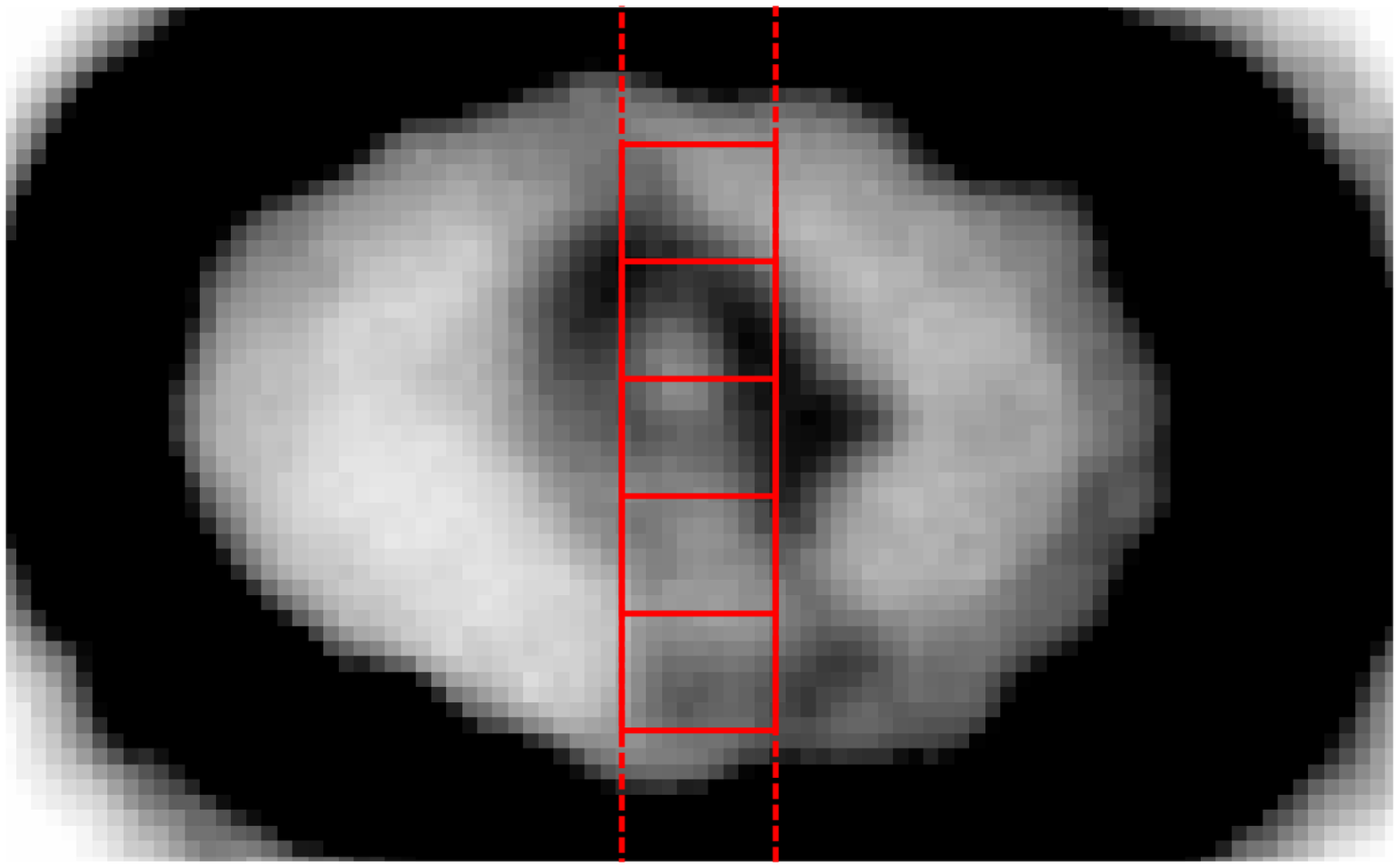}}
\resizebox{75mm}{!}{\includegraphics{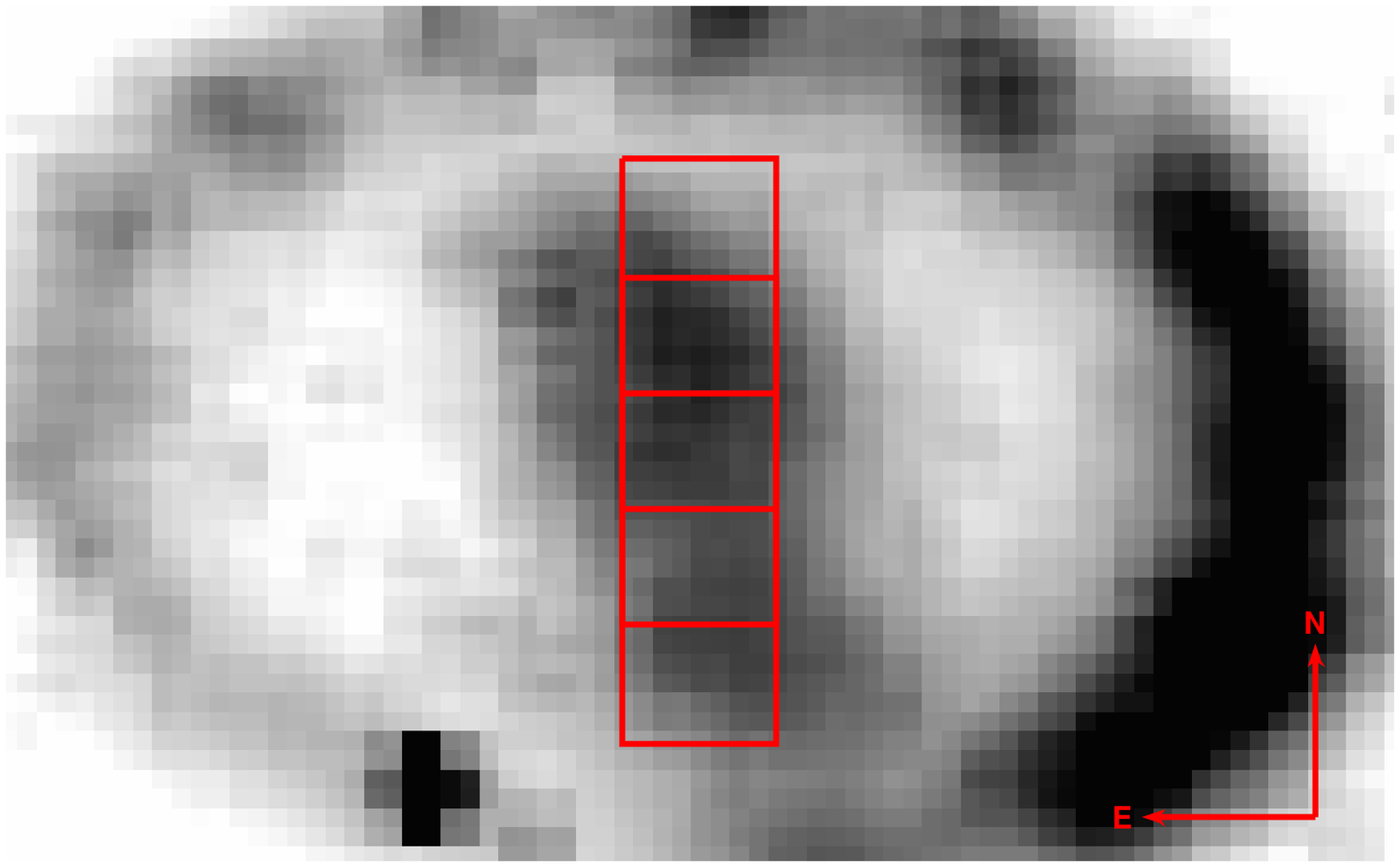}}
\caption{Extraction regions (solid red lines) used for producing the
  spectra in Figs.~\ref{spres_profiles05} (upper panels) and
  \ref{spres_profiles11} (lower panels). All the extraction regions are
  $0.2'' \times 0.15'' $ and the field of view for each of the images is $1.8\arcsec \times 1.1\arcsec$. Top left: The three slit positions used for
  the STIS observations on day 6,355 are shown as dashed lines
  superposed on the ACS F625W image from day 6,790. Top right: SINFONI
  image from day 6,816 in the wavelength range corresponding to
  1.644~$\mu$m $\pm 3,000~\rm{km}\ \rm{s}^{-1}$. Bottom left: The slit
  position used for the STIS observations on day 8,378 is shown as
  dashed lines superposed on the WFC3 F625W image from day 8,328. Bottom
  right: SINFONI image from day 8,714 in the wavelength range
  corresponding to 1.644~$\mu$m $\pm 3,000~\rm{km}\ \rm{s}^{-1}$.}
\label{regions0410}
\end{center}
\end{figure*}

\begin{figure*}
\begin{center}
\resizebox{\hsize}{!}{\includegraphics{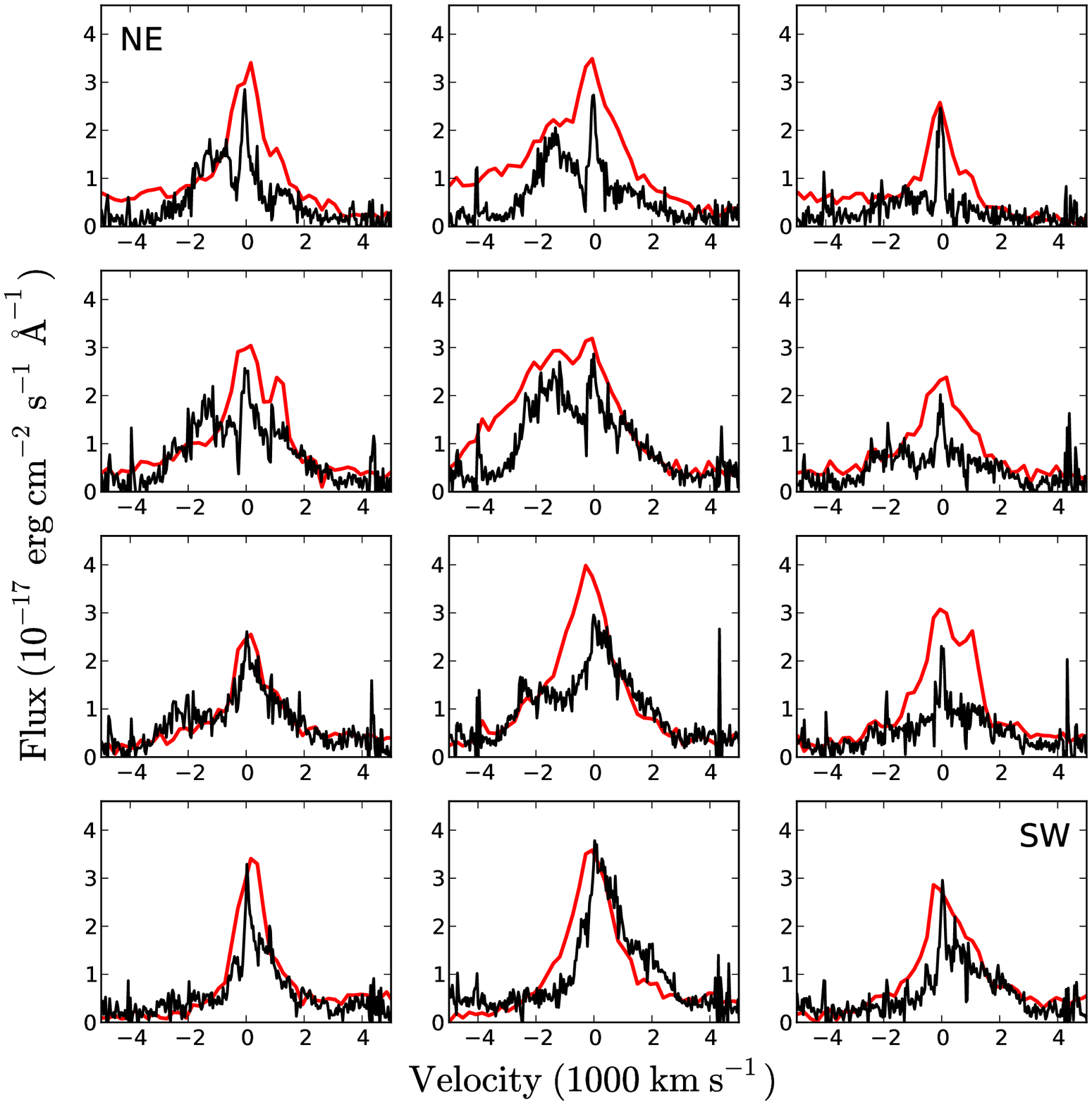}}
\caption{Spatially resolved STIS G750L (red) and SINFONI H-band
  (black) spectra from day 6,355 and 6,816, respectively. The STIS spectra
  are centered on zero velocity for H$\alpha$ while the SINFONI spectra
  are centered on zero velocity for the [Si~I]+[Fe~II]   line at 
  1.644~$\mu$m. The SINFONI spectra have been rescaled by a constant to
  match the flux level of the STIS spectra. Each spectrum was
  extracted from a $0.2\arcsec \times 0.15\arcsec$ region, as shown in
  Fig.~\ref{regions0410}.  
}
\label{spres_profiles05}
\end{center}
\end{figure*}

\begin{figure*}
\begin{center}
\resizebox{50mm}{!}{\includegraphics{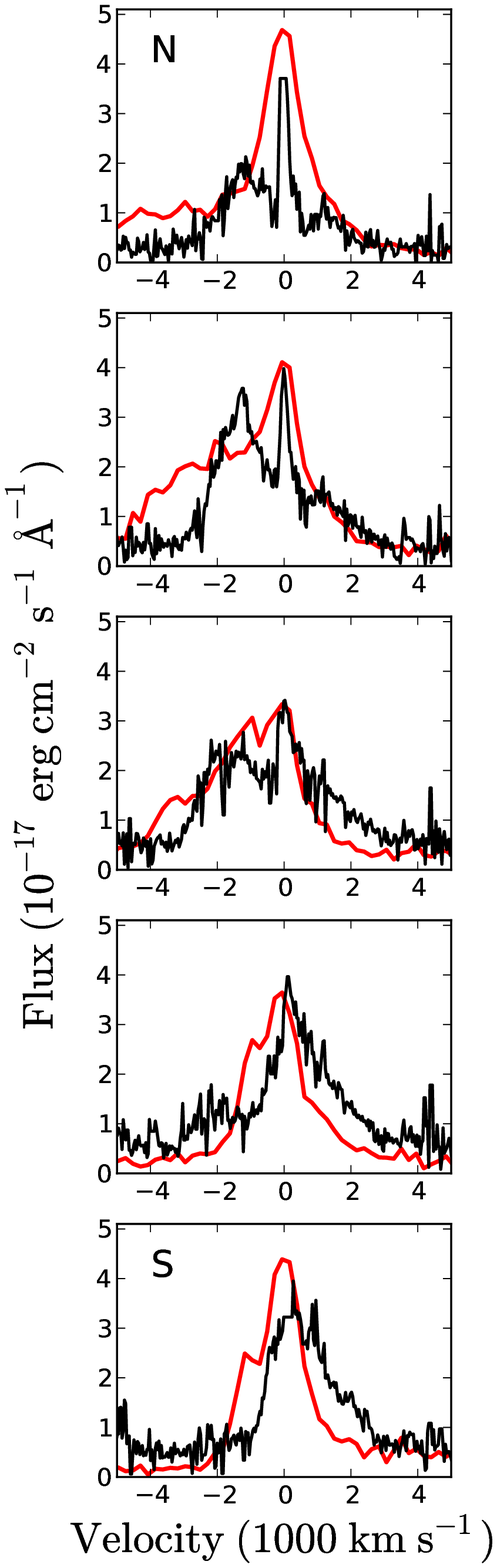}}
\caption{Spatially resolved STIS G750L (red) and SINFONI H-band
  (black) spectra from day 8,378 and 8,714, respectively. The STIS
  spectra are centered on zero velocity for H$\alpha$ while the SINFONI
  spectra are centered on zero velocity for the   [Si~I]+[Fe~II]   line at 1.644~$\mu$m. The SINFONI spectra have been rescaled by a
  constant to match the flux level of the STIS spectra. Each spectrum
  was extracted from a $0.2\arcsec \times 0.15\arcsec$ region, as shown in
  Fig.~\ref{regions0410}.}
\label{spres_profiles11}
\end{center}
\end{figure*}

\begin{figure*}
\begin{center}
\resizebox{75mm}{!}{\includegraphics{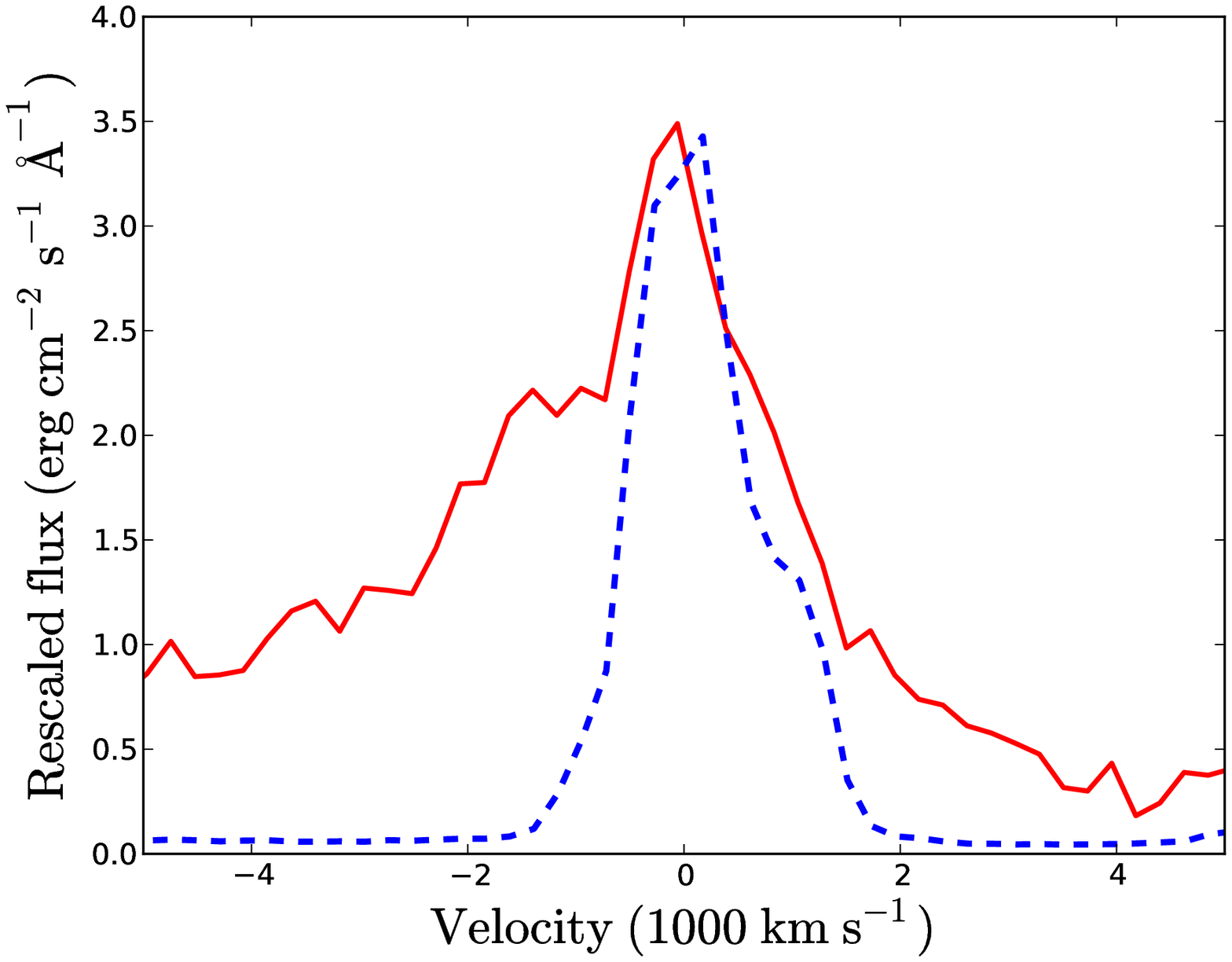}}
\resizebox{75mm}{!}{\includegraphics{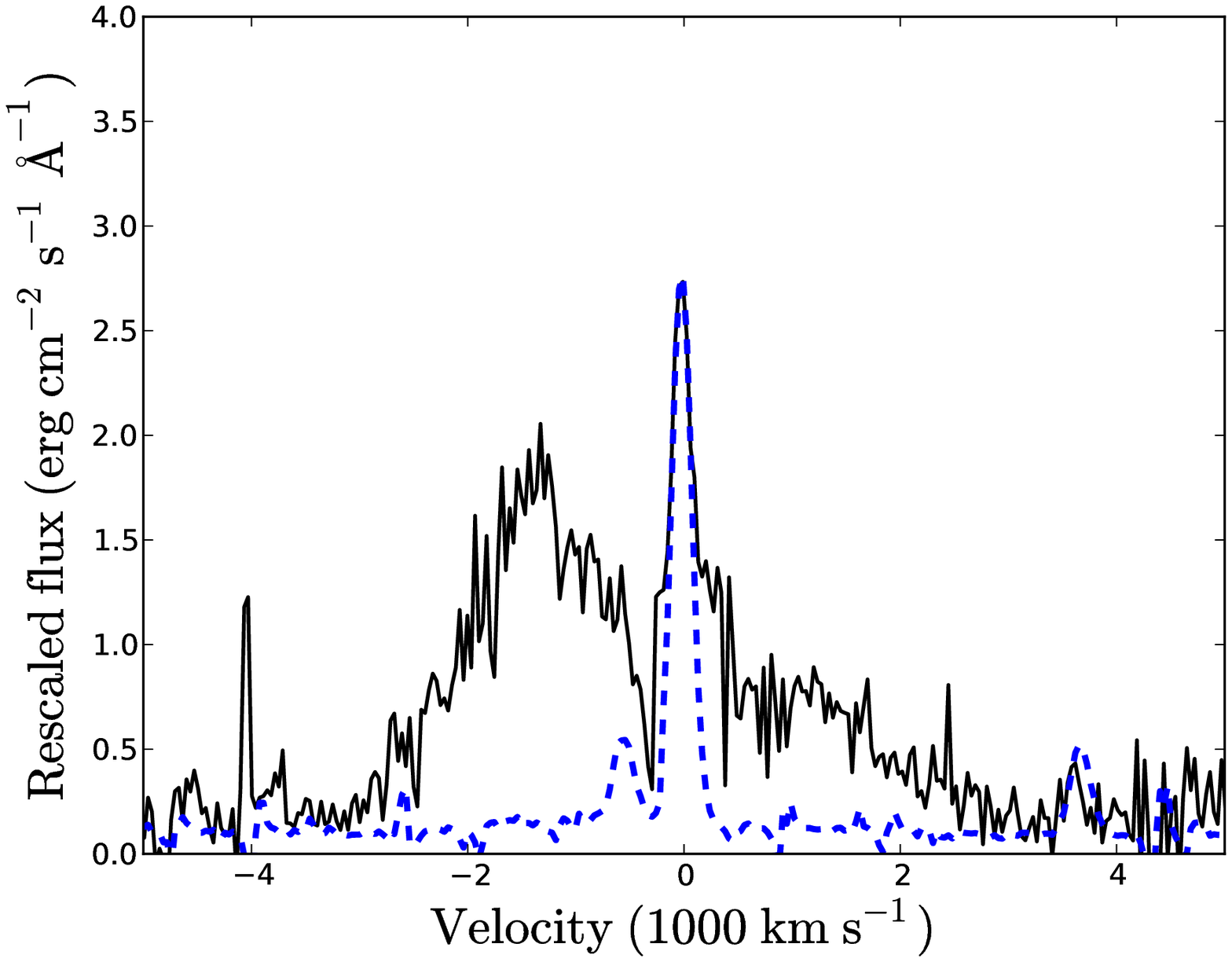}}
\caption{Ejecta spectra from the top, middle panel of
  Fig.~\ref{spres_profiles05} shown together with rescaled spectra of
  the equatorial ring. STIS spectra (day 6,355, centered on zero velocity for H$\alpha$) are shown to the left and
  SINFONI spectra (day 6,816, centered on zero velocity for the  1.644~$\mu$m [Si~I]+[Fe~II]   line) to the right. In both plots the scaled ring emission
  is shown as a dashed blue line, illustrating the maximum possible
  contamination from the ring. }
\label{ringcont}
\end{center}
\end{figure*}

In \cite{Larsson2011} we showed that the ejecta of \8 have brightened
by a factor 2-3 in the F675W/F625W and F439W bands since day 5,012. We
also showed that the brightening can be explained by a change in the
dominant energy source, from radioactive decay before day $\sim 5,500$
(year $2001$) to X-ray input from the ejecta - ring collision
thereafter. In this paper we have extended the analysis to a broader wavelength range,
using five broad filters that cover the entire interval
between $\sim2,000$ and $\sim10,000$ \AA.  As shown in  Fig.~\ref{sed},  the flux has increased
by a factor 2-3 between days 5,012 and 8,323 in all the filters, with
the exception of the UV-band (F255W/F225W), where the increase is a
factor of $\sim 5$.

There are no spectra in the UV-band at late epochs, so we cannot say
exactly which lines are responsible for the large increase, or if the flux correction between the F225W and F255W filters deduced from the early spectra has changed. However, we note
that at early epochs the band is dominated by Fe I-II and Mg I-II
lines (Fig.~\ref{throughput}) and that most of the flux in these lines
is due to resonance scattering in the H-rich zones of the
supernova \citep[section \ref{discussion_thermal}
  and][]{Jerkstrand2011}. As we will
discuss below, most of the X-rays are absorbed in the
H-rich outer ejecta, so it is reasonable that the largest increase is seen in this wavelength interval.
In fact, from figures 3 and 4 in \cite{Jerkstrand2011} it is clear that the core emits primarily in the $3,000-6,000$~\AA~range, while the envelope dominates the emission at shorter and longer wavelengths. This also agrees with the fact that the filter with the second to largest increase is F814W, at the longest wavelengths (Fig.~\ref{sed}). In this filter the dominant lines are the Ca [II] $\lambda \lambda 7291,7323$ doublet, the Ca II $\lambda \lambda 8600$ triplet and the Mg II $\lambda \lambda\ 9218.2, 9244.3 $ lines, all of which primarily come from the envelope. As discussed in F13, the Mg II lines may also be enhanced through Ly$\alpha$ fluorescence.

The evolution of the broad-band fluxes can be compared with the
spectral evolution determined from ground-based observations obtained
with UVES at the VLT (F13).  Although the spatial resolution is limited, the high
spectral resolution of the UVES data makes it possible to follow
several lines from the ejecta.  In particular, the broad emission lines from the inner ejecta can be separated from the narrow lines from the equatorial ring, as well as  the boxy, broad lines from the reverse shock. Using this method, F13 found an increase in the H$\alpha$ and [Ca II] $\lambda \lambda 7291,7323$ lines (which fall within
the F675W and F814W filters) by a factor of $4-6$ between 5,000 and 8,000
days. This is somewhat higher than we have found with HST (Fig.~\ref{sed}) and the difference gives an idea of the systematic uncertainties involved in the measurements. The main uncertainties are the contamination from the equatorial ring and differences in the fraction of ejecta contained within the slits and apertures used. The latter problem is enhanced by the fact that the ejecta brightness is highly nonuniform.  The comparison between the HST and UVES results are further complicated due to the fact that we are comparing broad-band fluxes with fluxes in individual lines. However, given that the continuum is rather low (Fig.~\ref{throughput}) this is unlikely to be the main reason for the differences.  
Due to all these systematic differences a direct comparison of absolute flux levels determined from HST and UVES is not meaningful.

The flux in the $1.644\ \mu$m line, which we measured from the SINFONI
observations on days 6,816 and 8,328, has also increased, but the rather
large uncertainties in the absolute flux calibration prevents us from drawing any strong conclusions from this.  As already noted above, it is possible that a significant fraction of the emission at these wavelengths is still due to radioactivity.

\subsection{Evolution of the morphology}
\label{discussion_hole}

\subsubsection{Edge-brightened morphology due to X-ray illumination?}

In this paper we have shown that the morphology of the ejecta seen in the HST images changes
from an elliptical shape to a more irregular, edge-brightened geometry
around the same time as the dominant energy source changes. A likely
explanation for the late-time morphology is thus that it is due to the
external X-ray illumination. The details of the X-ray deposition in
the ejecta are discussed in some detail in F13, so
here we only make some qualitative remarks that are relevant for the
morphology.

The location of  X-ray absorption in the ejecta depends on the X-ray spectrum. Specifically, the column density
penetrated by an X-ray photon with energy $E$ is $\propto E^{8/3}$. As
a result, the soft X-ray/extreme UV emission from the slow shocks in
the ring will be absorbed in the outermost parts of the ejecta, while the
hard emission ($\ga 2-3$ keV) can penetrate further in. 

As shown in \cite{Park2011} there is a considerable and increasing hard flux from the shocks
in the equatorial ring. Most of the hard emission is likely to be
absorbed at the boundary between the envelope and core region, where
the steep density gradient and the dramatic increase in metallicity
leads to an increase in the X-ray opacity \citep[][F13]{Larsson2011}. A central low surface brightness region,
largely protected from the X-ray illumination from the ring, is
therefore a natural outcome, as also demonstrated by the modeling in F13.

In view of this model it is also interesting to consider the top panel of Fig.~\ref{shrunk_contours}, which shows the temporal evolution of the R-band morphology together with a circle that follows the homologous expansion. In most of the ejecta the emission follows the homologous flow.  This is expected in both a radioactively- (or freeze-out-) dominated scenario, as well as in the case where most of the X-rays are thermalized in the steep density gradient between the envelope and the core. However, the southern part of the ejecta clearly 'expands' faster. This could indicate a dense region moving at high velocities which was not reached by the early radioactive input, but which is now being illuminated by the X-rays. 

Looking in more detail at the late-time morphology in the different
bands in Fig. \ref{filterim}, we see that the emission
appears somewhat more spatially extended in the F225W and F814W
filters than in the other bands. In the former case this is likely due
to the resonance scattering discussed in section \ref{discussion_thermal}. The forest of
strong lines in the 2000-3000 \AA \ region have large optical depths
and may scatter emission at lower densities and thus larger radii than
in the redder bands. 

In the case of the F814W band the emission is
dominated by the Ca [II] $\lambda \lambda 7291,7323$ doublet, the Ca II $\lambda \lambda 8600$ triplet and the Mg II $\lambda \lambda\ 9218.2,
9244.3 $ lines. The Mg II
emission is likely to be a result of Ly$\alpha$ fluorescence in the hydrogen
envelope (see discussion in F13), which may result in a more extended emissivity. However,
confirmation of these issues will require a detailed spectral modeling,  
including the X-ray input.

One possible explanation for the more centrally peaked morphology in the SINFONI images is that we are still probing the radioactive input with these observations.  However, any 
hard X-ray emission penetrating the inner core of the ejecta would also result in stronger [Si I] and [Fe II]  emission from this region compared to H$\alpha$. The reason for this is that the non-thermal excitation by the electrons produced in the X-ray photoionization is roughly  proportional to the abundance (in contrast to
thermal excitation, which is mainly temperature-dependent) and the abundance of Si and Fe is higher in the core.  

In this context it is also important to point out why the morphology
seen by SINFONI differs from the morphology in e.g.~the F225W/F255W
band, which is also dominated by metal lines. As discussed 
above, the emission in this band is dominated by resonance scattering in the
 H-rich gas. A large fraction of the UV/optical
emission from the metal core will thus have its last scattering in the
hydrogen envelope, while the emission at near-IR and longer wavelengths
can escape directly from the core.

\subsubsection{Correlation between ring and ejecta emission}

If the ejecta are powered by radiation emanating from the
equatorial ring, we would expect some degree of spatial correlation
between the flux from the ring and the ejecta (we have already shown
that the total fluxes correlate, \citealt{Larsson2011}). The $\sim 6$
month light travel time between the ring and the inner ejecta is,
however, too short to probe with the available observations. 

The first optical hot-spot appeared in the north-east part of the ring
\citep{Sonneborn1998,Lawrence2000}, and a corresponding brightening
can indeed be seen in the closest part of the ejecta (days 5,012 and
6,122 in Fig.~\ref{contours}). As more and more optical hot-spots
appeared a larger fraction of the ejecta started brightening, as is
apparent from Fig.~\ref{contours}. A caveat here is of course that
the hot-spots observed in the optical/UV are produced by radiative
shocks with velocities $\la 500\ \kms$ \citep{Groningsson2008b}. They
therefore mainly reflect the softest X-rays with energies $\la 0.3$
keV, which are expected to be absorbed in the outermost parts of the
ejecta. The resulting effect on the morphology is difficult to judge
due to the low surface brightness of this region. Slow recombination (freeze-out) may also
delay the optical ejecta emission compared to the X-rays from the ring.

For the harder X-rays, which are more important for the deposition
further in, we have to compare with the Chandra images. A set of 0.3 -
8~keV images from day 4,711 to day 7,987 are shown in \cite{Park2007}
and \cite{Racusin2009}. Unfortunately, no separate images are shown
for the different X-ray energies. While Chandra observations
at around 5,000 days showed a different morphology above and below 1.2
keV \citep{Park2002},   \cite{Park2007} state that the
later images up to 2007 show a similar morphology in the hard and soft
X-ray bands.

Although the resolution of the Chandra images is considerably lower
($\sim 0.5 \arcsec$) than the HST images, we note that the brightening
in the X-rays first occurred in the north-east and south-east parts of the ring
\citep[see Fig. 1 of ][]{Park2007}. In particular, the north-east brightening
corresponds to a similar brightening of the ejecta in the HST image at
5,012 days (Fig.  \ref{contours}). The south-east X-ray brightening does not,
however, have a prominent correspondence in the ejecta.  Possible reasons for this are noted below. The next region to brighten
in the Chandra images is the north-west part, which increases dramatically
between day 5,791 and day 6,533.  Again, a similar evolution can be seen
in the day 6,122 and day 7,226 ejecta images. The last Chandra region to
turn on is the south-west, which is also followed by a brightening in the HST
ejecta images after day 7,226.

To summarize, we find a fairly good correspondence between the X-rays
emission from the ring and the HST images of the ejecta, as expected
from the X-ray dominated scenario for the energy input. A more
detailed analysis is beyond the scope of this paper, as it would have
to take into account the difference in the azimuthal X-ray
illumination between the soft and hard bands, the
non-spherical distribution of the ejecta, as well as the possibility of dust obscuration.

\subsubsection{Effects of the Ni-bubble and dust}
\label{dust}

Although it is clear that the X-ray illumination can explain the edge-brightened morphology in the HST images, it is also possible that the observed morphology is affected by the 'Ni-bubble' or the presence of dust in the ejecta. We now discuss these alternatives in turn.

The Ni-bubble is expected to occur as a result of heating by the
${}^{56}$Ni decay during the first days after the explosion
\citep[e.g.][]{Herant1992,Li1993, Ellinger2012}. The bubble (which may be composed of many small bubbles, \citealt{Li1993}) has
the effect of pushing the nuclear burning zones located outside the
Fe core outwards, thus creating a dense shell separating the
H- to O-rich zones from the Fe-rich material. This could
in principle explain the difference in morphology between the SINFONI
and HST images (Fig. \ref{comp_contours}). 

However, we believe that this explanation is unlikely since we find
that the ejecta morphology has changed considerably over the observed time
interval, in contradiction with the expectation that the Ni-bubble
should be frozen into the homologous expansion. An additional problem
with this explanation is that the filling factor of the Ni-bubble was
estimated to be only $\sim 0.2$ \citep{Kozma1998} within $\sim
1,800\ \kms$, which is likely to be too small to explain the
observations.  The filling factor estimated by \cite{Kozma1998} is smaller than the results of \cite{Li1993} (who find a filling factor of $\sim 0.6$), but the latter work assumed that no primordial Fe contributes to the line emission.

There is plenty of evidence for the formation of dust in SN~1987A, both from line profiles \citep{Lucy1989} and thermal emission during the
first years \citep{Wooden1993}, as well as recent Herschel \citep{Matsuura2011}
 and sub-millimeter \citep{Lakicevic2011} observations, although the amount of dust formed is highly
debated. Any dust present in the ejecta is likely to have formed in the
metal-rich core of the SN, and could thus cause an obscuration of the
central regions. The difference between the SINFONI and HST images
could then partly be an effect of the lower extinction in the near-IR SINFONI
band. 

In order for this effect to be important the dust has to be optically thick in the F814W filter at $\sim 0.85\ \mu$m
(where the hole is seen, Fig. \ref{filterim}), but not in the SINFONI  H-band at $\sim 1.6\ \mu$m.  The comparatively small difference in extinction
between 0.85 $\mu$m and 1.6 $\mu$m (a factor of $\sim 3$) then
requires that we observe at just the right epoch when these conditions
are met. Since the optical depth of the dust is expected to decrease
as $t^{-2}$, where $t$ is the time since explosion, substantial fine
tuning is required in order for optical depth unity to be reached at
$\sim 20$ years. In addition, the decrease in optical depth would be
considerable during the time covered by our observations, leading to an increased transparency at
H$\alpha$, which is not indicated in the HST images.

It is more likely that the dust is in the form of very optically thick clumps, which would give the same extinction at optical and infrared wavelengths. This is supported by the fact that the [O~I] $\lambda \lambda$ 6300, 6364 lines, which became asymmetric after about 500 days, presumably as a result of dust formation \citep{Lucy1989} still have very similar profiles around  4,000 days (F13). These line profiles, as well as the Mg I] $\lambda$~4571 line \citep{Jerkstrand2011} indicate an effective optical depth close to unity, which in the case of optically thick clumps represents a covering fraction of the clumps.

Absorption by dust in the ejecta will primarily have the effect of reducing the redshifted emission. As clearly seen in Fig.~\ref{spres_profiles05}, the red halves of the profiles decay smoothly (and in a similar way for H$\alpha$ and [Si~I]+[Fe~II]) while the blue profiles rise more sharply for [Si~I]+[Fe~II]  . This indicates that Si/Fe is distributed in more sharply defined clumps than H, but that both components are affected by dust absorption.  More definite conclusions regarding the effects of dust on the observed morphology and line profiles will have to await spatially resolved images of the dust emission in the ejecta. This will hopefully be achieved with ALMA in the near future.

\subsection{Evolution of the line profiles}

Our study of the ejecta spectra at different epochs does not reveal
any strong temporal evolution of the line profiles.  In particular, we do not see any major changes between the H$\alpha$ profiles from day 4,571 (when radioactive decay was still
dominating the energy input, Figs.~\ref{spres_profiles99m} and \ref{full_profiles_stis}) and  day  6,355 (when the X-ray input had started to dominate, Figs.~\ref{spres_profiles05} and \ref{full_profiles_stis}). We note, however, that the comparison is made difficult by the change of grating, slit width and slit orientation.

The change in the dominant energy source is not expected to produce any sudden, dramatic changes in the spectra. Both the
radioactive decay and the X-ray emission result in secondary
non-thermal electrons, which subsequently give rise to line emission
via the same physical processes, thus producing similar spectra. (see
section \ref{discussion_thermal} above, \citealt{Larsson2011} and
F13). In addition, the freeze-out effect in the
envelope will delay changes in the energy input. The main difference
between the two phases is that the energy deposition occurs at higher
ejecta velocities in the X-ray dominated phase. The location of
the X-ray deposition is also likely to change with time as the ejecta expand and the
X-ray light curve evolves. These changes should be reflected in the observed line profiles.

No major changes have been seen yet, as evident from a comparison of the spatially resolved profiles in Figs.~\ref{spres_profiles05} and \ref{spres_profiles11}, which show
the H$\alpha$ and [Si~I]+[Fe~II]   emission around $6,600$ and $8,500$
days, respectively.  However, the one slit that was
used in the latter STIS observation does not contain the blob in the
western part of the ejecta, where the strongest brightening is seen in
the HST images.

An important lesson from this work is that the total line profiles of
the ejecta are only slightly asymmetric
(Fig.~\ref{full_profiles_stis}), despite the large asymmetries seen in
the images and in the spatially resolved line profiles. The fact that
such prominent asymmetries can nearly cancel out in the integrated
profile highlights the uncertainties involved in trying to obtain
information about the explosion geometry from the spectra of distant
supernovae, where no spatial information is available.

\subsection{Spatial distribution of the ejecta and implications for explosion models}
\label{discussion_explosion}

\begin{figure*}
\begin{center}
\resizebox{53mm}{!}{\includegraphics{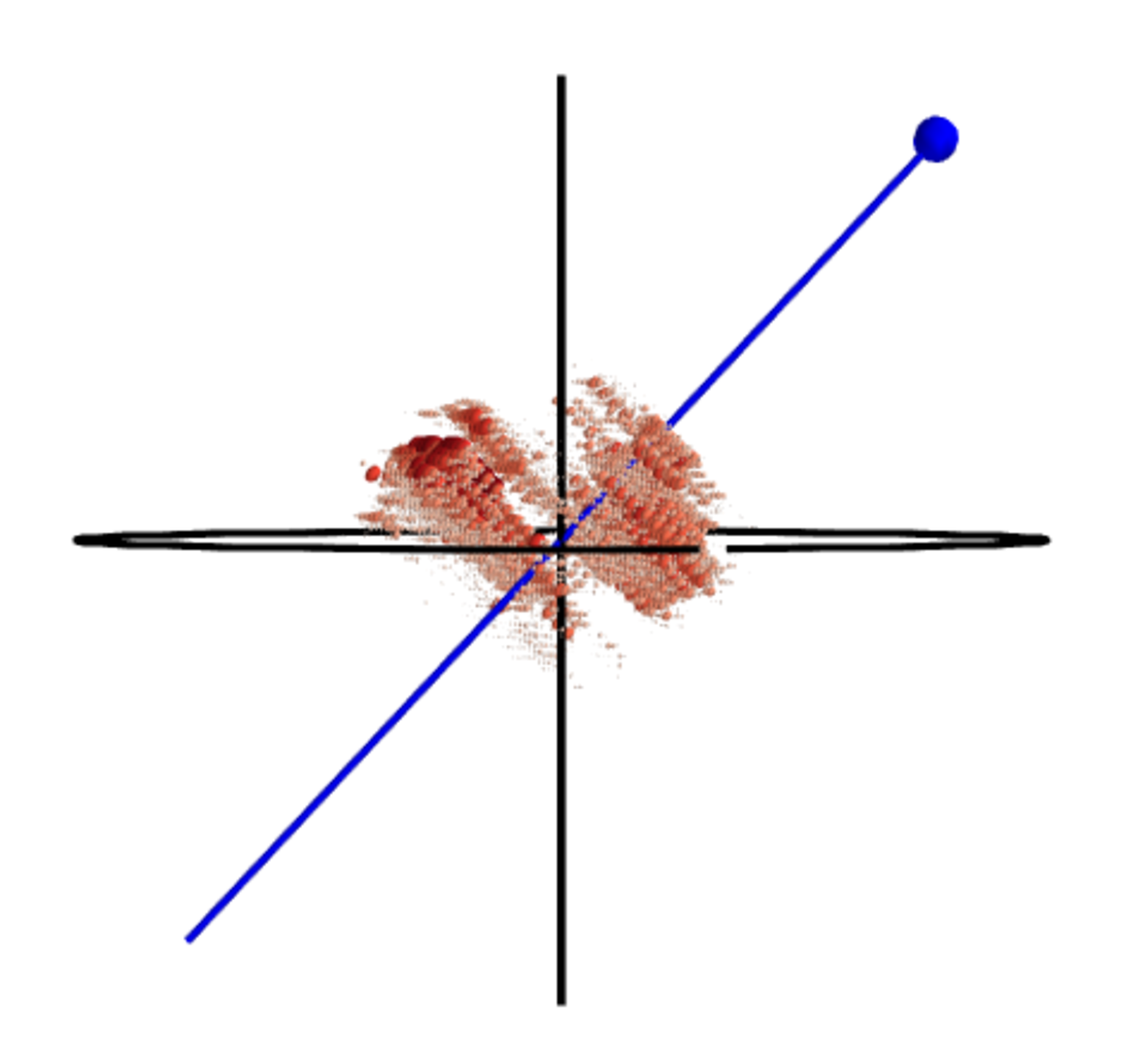}}
\resizebox{53mm}{!}{\includegraphics{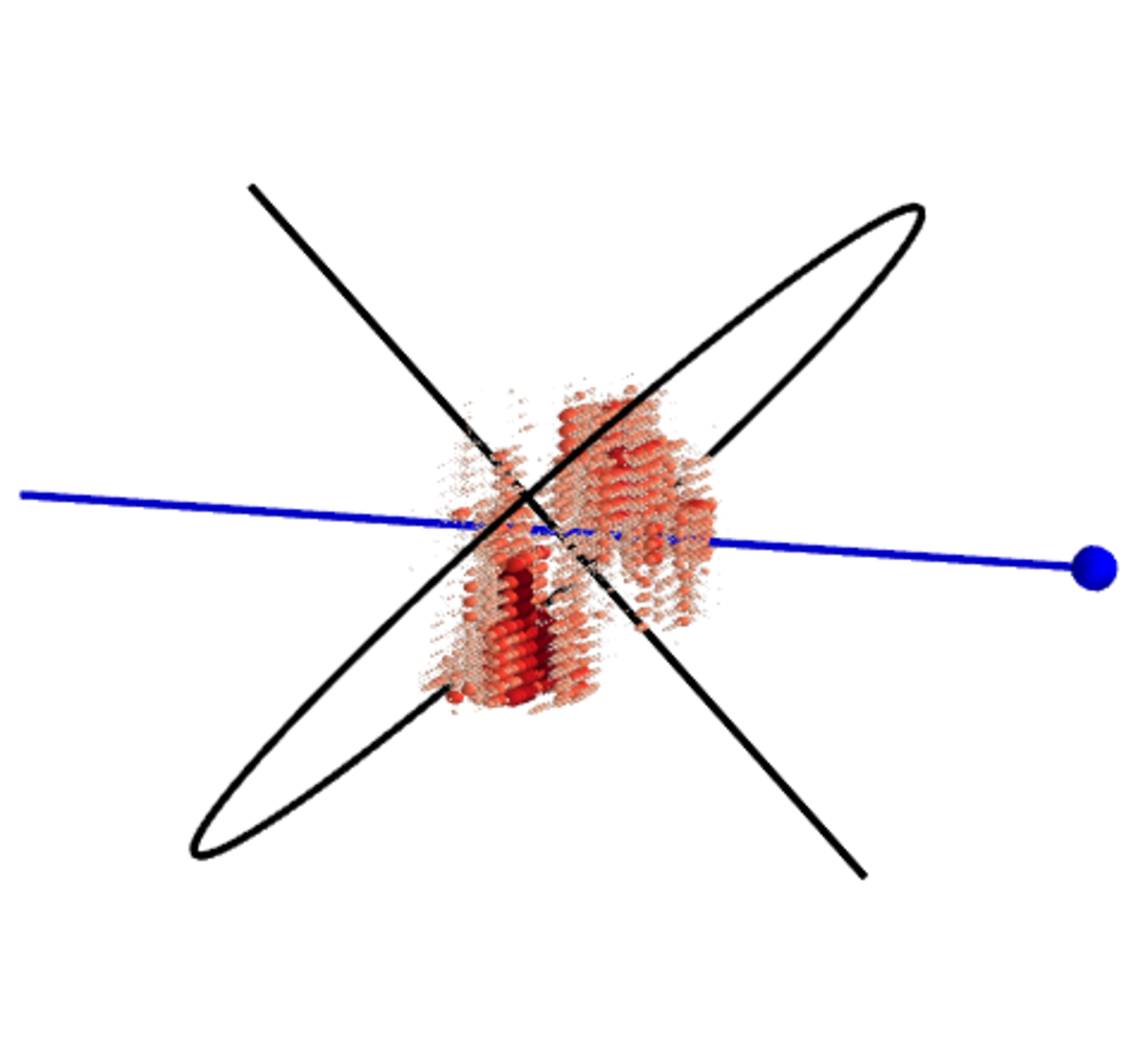}}
\resizebox{53mm}{!}{\includegraphics{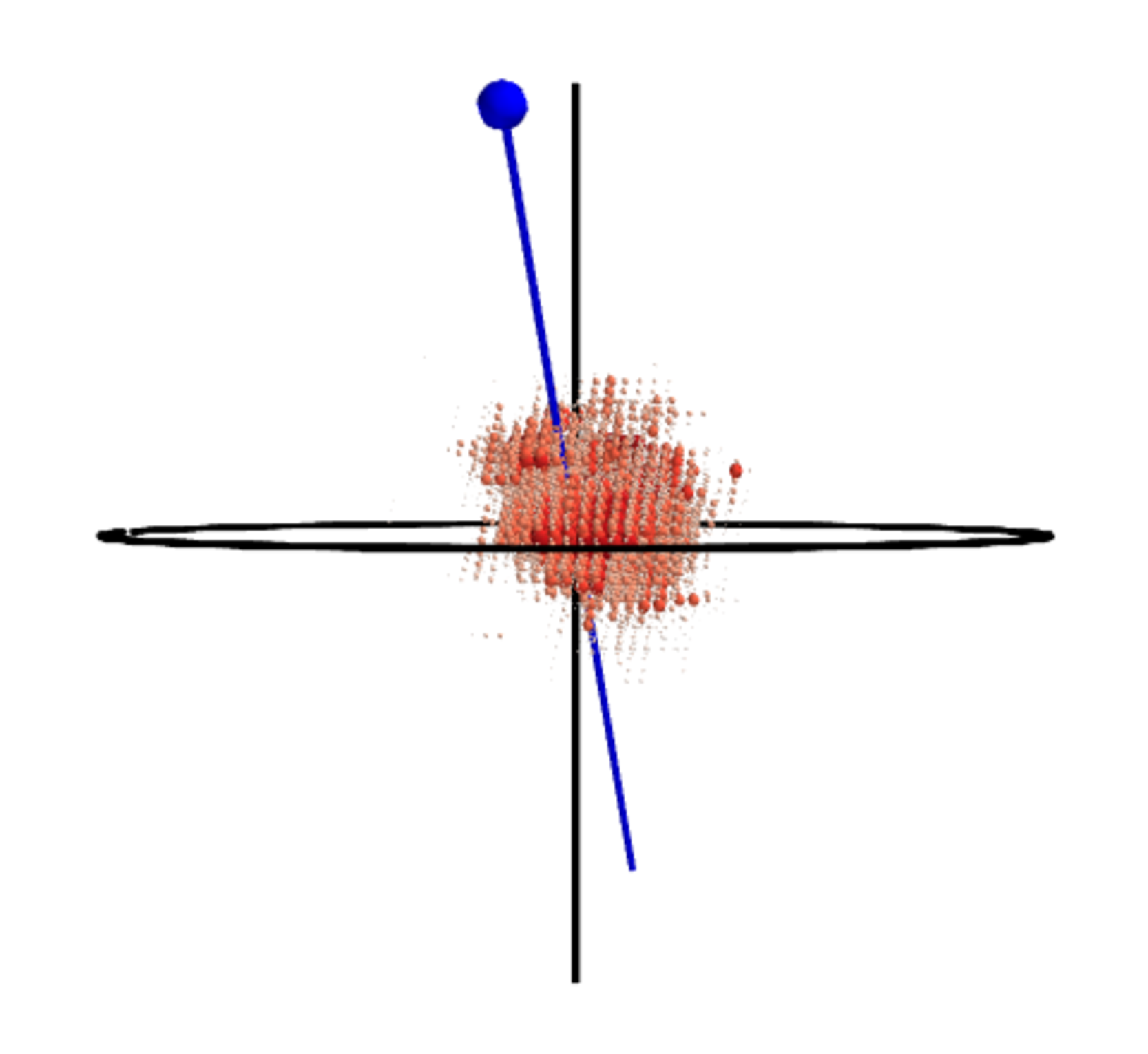}}
\resizebox{53mm}{!}{\includegraphics{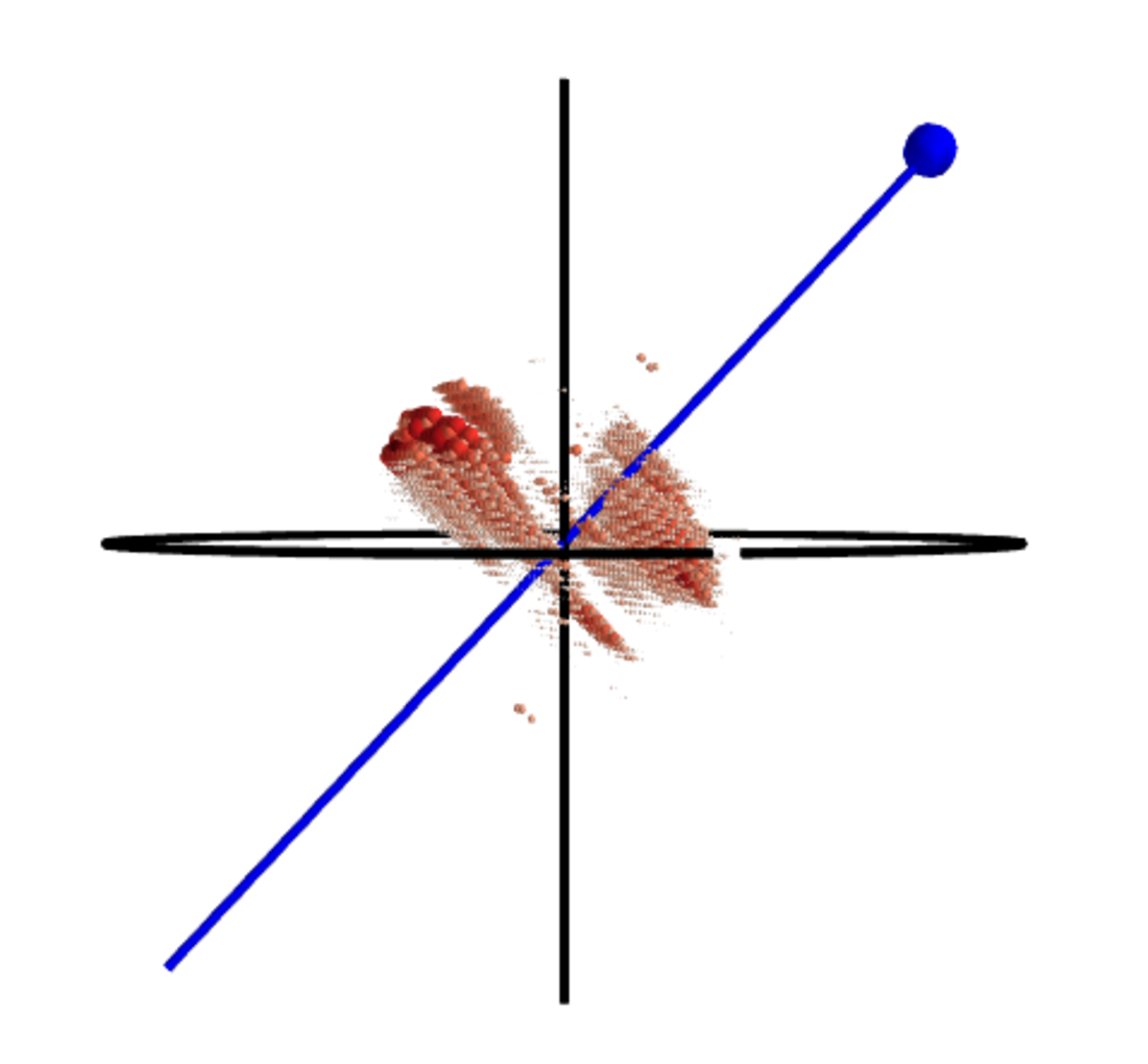}}
\resizebox{53mm}{!}{\includegraphics{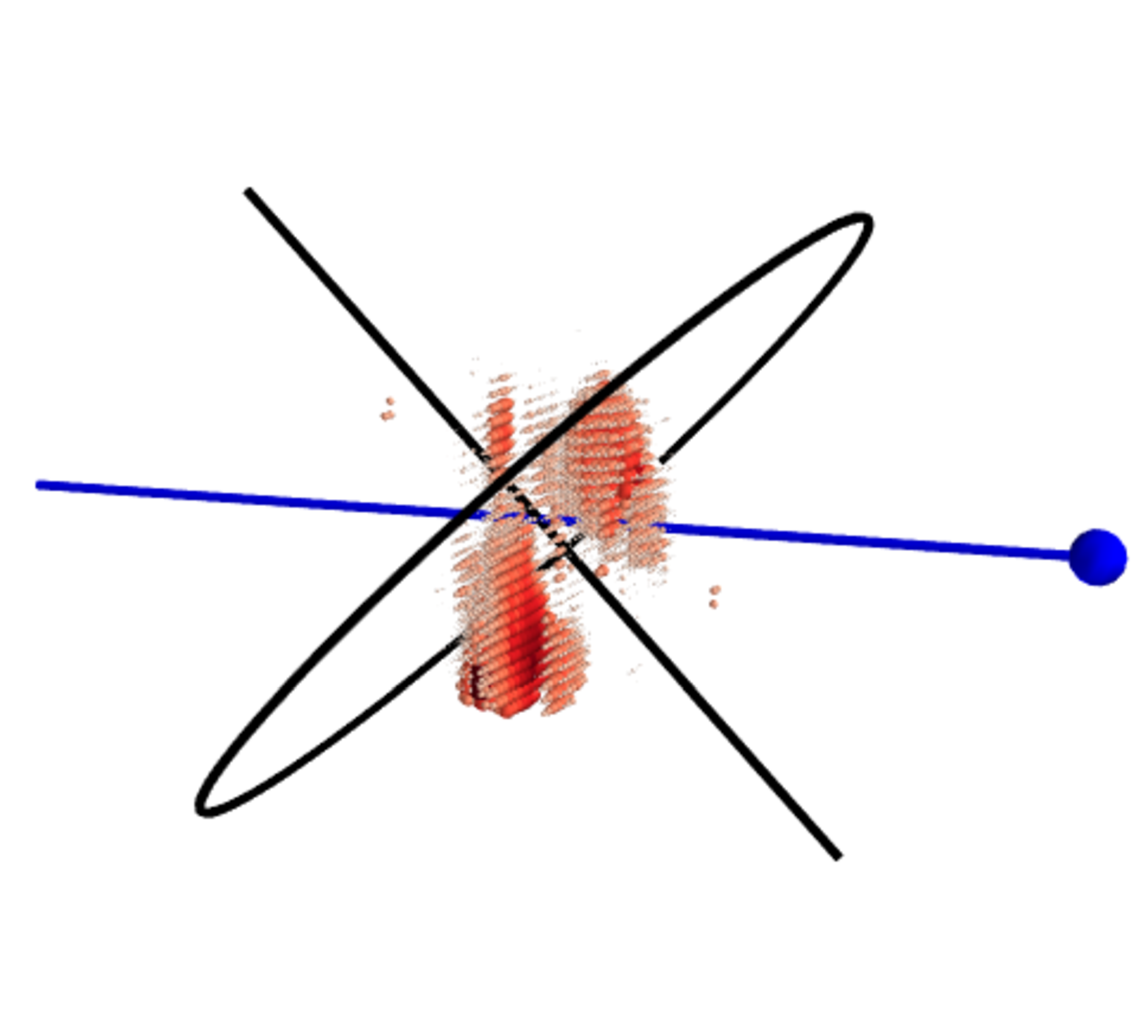}}
\resizebox{53mm}{!}{\includegraphics{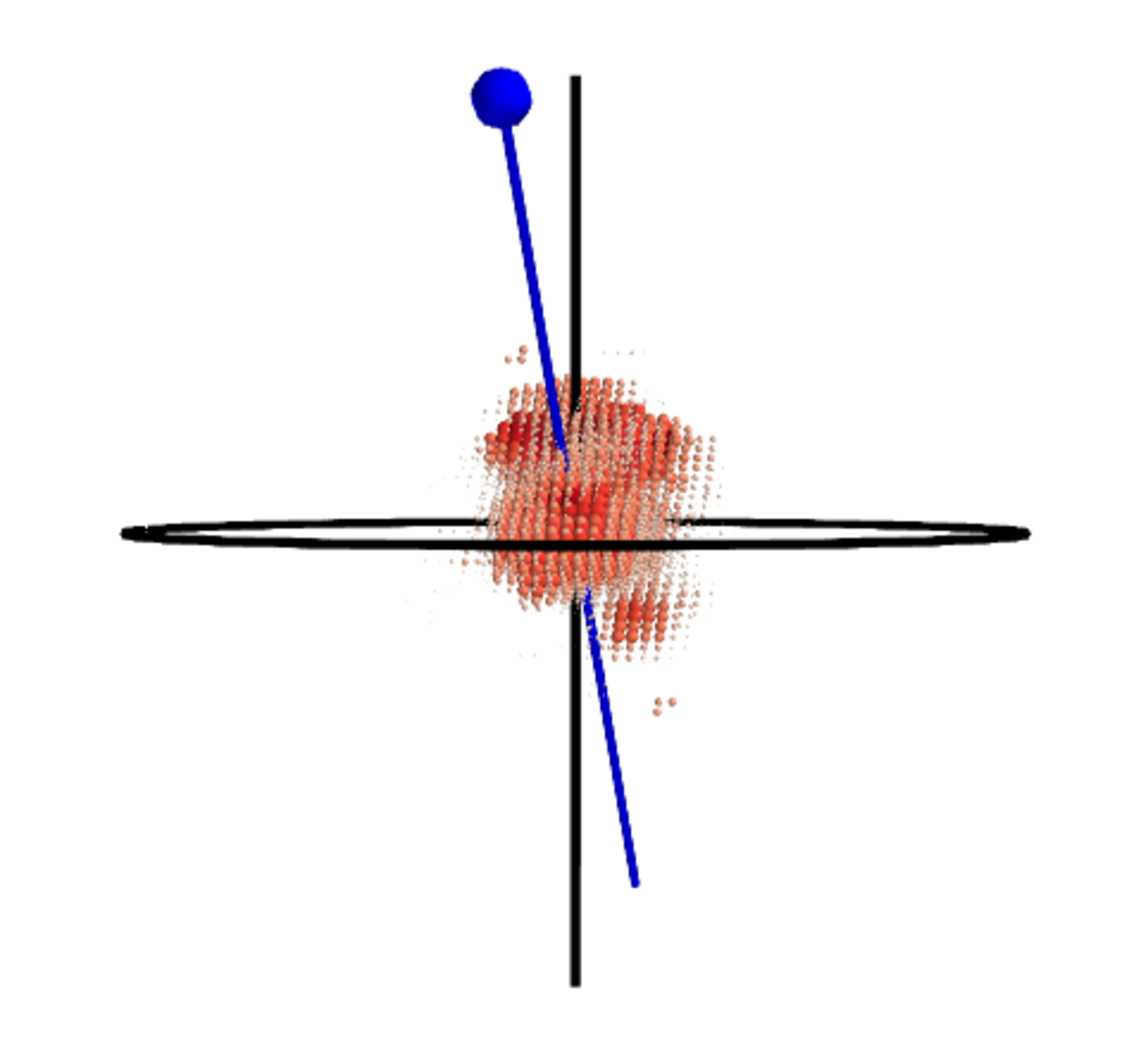}}
\caption{Three-dimensional view of the 1.644~$\mu$m emission from the
  ejecta observed by SINFONI on day 6,816 (top row) and 8,714 (bottom
  row) for different viewing angles. The blue sphere and line show the
  position of the observer and the line of sight. The black circle
  shows the peak of the emission from the equatorial ring, and the
  black line is perpendicular to the plane of the ring. The radius of the ring is $0.82\arcsec$, which corresponds to $6.1 \times 10^{17}~\rm{cm}$ at a distance of 50~kpc. The
  plots show ejecta emission brighter than 3 times the continuum level
  for each epoch, with darker and bigger spheres representing brighter
  emission. The ejecta emission is plotted out to $\pm
  3,500~\rm{km\ s^{-1}}$ (assuming homologous expansion), excluding the interval between $\pm
  250~\rm{km\ s^{-1}}$ along the line of sight, which is contaminated by emission from the
  ring.  Animations of the 3D emissivity are available in the online journal.}
\label{sinf3d}
\end{center}
\end{figure*}

\begin{figure*}
\begin{center}
\resizebox{80mm}{!}{\includegraphics{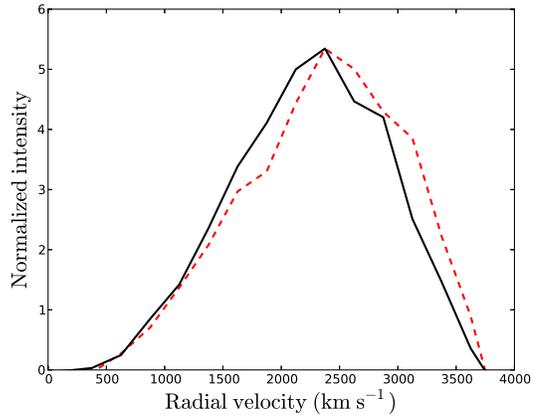}}
\caption{Radial distribution of the [Si~I]+[Fe~II]    1.644~$\mu$m emission from the
  SINFONI observations on day 6,816 (red dashed line) and 8,714 (black
  line), summed over spherical bins with velocity width
  $250\ \kms$. The continuum was calculated over the $3,500 -
  4,000\ \kms$ interval and subtracted from the other
  bins. }
\label{radial_dist_sinf}
\end{center}
\end{figure*}

All the observations that we have considered in this paper show that the ejecta morphology is highly non-spherical. There is no well-defined symmetry axis for all the emission, but it is clear from Fig.~\ref{sinf3d} that the [SiI]+[Fe II] emission is closer to the plane of the ring than perpendicular to it (as already noted by \citealt{Kjaer2010}).  The H$\alpha$ emission is consistent with having the same
distribution relative to the ring plane, but the emission  extends to higher velocities than [Si~I]+[Fe~II].  The fact that the ejecta are preferentially distributed close to the plane of the ring has also been inferred from observations of light echoes \citep{Sinnott2012}, although it should be noted that these observations primarily probe the outermost ejecta.

We also found that there is a north-south asymmetry for both  [Si~I]+[Fe~II]    and H$\alpha$;  the strongest emission in the southern part is concentrated closer to the plane of the sky, while the strongest emission in the northern part is closer to the plane  of the ring.  As discussed above, this asymmetry may be enhanced by dust absorption.

One of the main motivations for this study was to obtain information
 about the explosion geometry. As discussed in
 section \ref{discussion_thermal} above, the observed emissivity distribution
 does not necessarily trace the matter distribution, but is a complex
 convolution of the input from radioactivity or X-rays, the density
 and abundance distribution, dust absorption, as well as the thermalization process.

In the case of [Si~I]+[Fe~II]   we are clearly seeing the innermost regions of the ejecta, and it therefore seems likely that the observed emission is a  reasonably good representation of the density distribution, regardless of what the dominant energy source is (see section \ref{discussion_hole}). In the case of H it is harder to relate the observed morphology to the density distribution since the morphology changes with time, responding to changes in the energy source. However, for the X-ray dominated scenario we have argued that the hole and the ringlike structure can be interpreted in terms of the high-density core and the hydrogen envelope just outside the core, respectively. This means that the late-time observations still provide useful information about the structure of the ejecta, even though the emission is not directly proportional to the density.  As discussed above, dust absorption is unlikely to significantly affect the observed morphology in other ways than reducing the intensity of the redshifted emission. 

  
  Keeping the above caveats in mind we compare the observed morphology with existing multidimensional simulations of the
 explosion geometry. In the absence of detailed predictions for the
 H$\alpha$ and [Si~I]+[Fe~II]   intensity distributions from these
 models we limit ourselves to some qualitative comparisons. The most detailed calculations of the explosion geometry are the 3D
 simulations by \cite{Hammer2010}. These calculations start from a
 self-consistent explosion model which provides the initial
 non-spherical irregularities that act as seeds for the continued
 evolution. Unfortunately, the simulations do not include the heating
  by the ${}^{56}$Ni decay, which creates the 'Ni-bubble'. They are
 also based on a 15 $\msun$ progenitor, which is at the lower end of the published mass estimates for \8. \cite{Fransson2002} find a progenitor mass of  $18-20 \msun$  based on the estimated O mass, while \cite{Smartt2009} estimate a progenitor mass in the range $14 -20 \msun$.

 Compared to similar, previous calculations in \cite{Kifonidis2006},
 these 3D calculations are found to give considerably more efficient
 velocity mixing of the different elements, including Ni-blobs with
 high velocities. These fast clumps have also
 been seen in later SPH simulations by \cite{Ellinger2012}. Even
 though \cite{Ellinger2012} do not start from a self-consistent explosion
 model, they do include the ${}^{56}$Ni decay, and also run their
 simulations for a much longer time than \cite{Hammer2010}, making
 sure that the effects of the radioactive heating are included.


 Starting with the [Si~I]+[Fe~II]   emission, we note two prominent
 blobs in the 3D maps shown in Fig. \ref{sinf3d}. The reduction in
 flux at low velocities seen in this plot is  exaggerated
 because the $\pm 250\ \kms$ interval along the line of sight was
 removed due to the strong contamination from the equatorial ring. However, a
 real decrease at low velocities is present, clearly seen in e.g. the
 dips on both sides of the narrow line from the ring in the northern
 part of the ejecta (Figs.~\ref{spres_profiles05} and
 \ref{spres_profiles11}). These clumps can be compared with the
 \cite{Hammer2010} simulations (their Figs. 2 and 3), where some
 prominent 'Ni' clumps (which include both Si and Fe) extend out to high velocities.  However, despite the qualitative agreement, our
 observations show a significantly  higher degree of asymmetry on large scales.
 Our integrated radial distribution of emission
 (Fig.~\ref{radial_dist_sinf}) also peaks at a higher velocity ($\sim
 2,300\ \kms$) compared to the radial mass distribution of 'Ni' in
 Fig.~6 of \cite{Hammer2010}($\sim 1,200\ \kms$).


The observations of H$\alpha$ in \8 have lower spectral resolution and are also harder to relate to the full density distribution. We therefore limit our comparison with explosion models to two very general properties of the observed emission. These are that the H$\alpha$ emission extends to higher velocity than [Si~I]+[Fe~II]  and that the the emission decays more smoothly with velocity than [Si~I]+[Fe~II]. Both of these conclusions are in qualitative agreement with the results of \cite{Hammer2010}. 

From the late-time HST images we can also estimate the size
scale of different structures in the ejecta, bearing in mind that this
only gives projected sizes and that we cannot resolve structures smaller than $\sim 500\ \kms$. The diameter of the 'hole' is around
$1,000 - 2,000\ \kms$ for all the late-time images, but the center
of the hole is offset by $\sim 800\ \kms$ from the center of the
remnant. The width of the ring-like structure corresponds to a
velocity width between $\sim 1,000 - 3,500\ \kms$. As noted above, we may identify the faint, central regions with the
high-density core and the ring structure with the hydrogen envelope
just outside the core. The brightest emission in the SINFONI images
is close to the faintest regions in the HST images, in agreement
with this interpretation.


In terms of studies of other young supernova remnants the most detailed results regarding the ejecta structure have been obtained in the case of Cas A  \citep{Fesen2006,DeLaney2010,Hwang2012}. A comparison with our results is not straightforward since both the supernova type and the type of observational data available are different.  In particular, the fact that Cas~A was a type IIb SN \citep{Krause2008} means that the H mass was very small; something which will strongly affect both expansion velocities and instabilities \citep{Iwamoto1997}. Compared to \8, the Cas A observations have dramatically higher spatial resolution, and also differ from our observations in that most of the observed emission comes from ejecta passing the reverse shock. Despite these differences, it is interesting to note that the knots, jets and pistons in Cas A have been suggested to have a flattened distribution \citep{DeLaney2010}.  While we do not detect the same type of structures, we find that the ejecta in \8 also have a flattened distribution.


In summary, we find reasonable qualitative agreement with the models,
but we stress that a more detailed comparison requires both more
realistic spectral modeling, as well as explosion simulations with
parameters matching SN 1987A and all physical effects included. In
terms of observations,  future ALMA observations of the
spatial distribution of dust and molecules will help reduce uncertainties about the
three-dimensional ejecta distribution. HST STIS observations with
narrow slits covering the full ejecta would also greatly improve the
comparison between the [Si~I]+[Fe~II]   and H$\alpha$ emission.

\section{Conclusions}

We have studied the morphology of the ejecta of \8 using HST images
obtained between day 2,770 and 8,328. In addition, we have obtained information about the 3D geometry from HST STIS spectra (for three
epochs after 4,500 days) and SINFONI integral field spectroscopy
 (for two epochs after 6,800 days). The HST imaging covers the $2,000 -
10,000\ \rm{\AA}$ interval, while the STIS spectra mainly probe the H$\alpha$
emission. For the SINFONI data we focused on the $1.644~\mu$m [Si
  I]/[Fe II] emission. Our main conclusions are as follows:

\begin{itemize}

\item{There is a strong temporal evolution of the morphology in the
  HST images, from a roughly elliptical shape before $\sim 5,000$ days,
  to a more irregular, edge-brightened morphology (with a 'hole' in
  the middle) thereafter.  The transition is most likely due to the
  fact that X-ray emission from the equatorial ring starts to dominate
  the energy input to the ejecta at this point, with most of the X-rays being absorbed at the boundary between the envelope and core region.   We find that other effects, from either the Ni-bubble
  or dust, are unlikely to significantly effect the morphology. 
  The X-ray input also results in a brightening
  of the ejecta, as we have previously shown in \cite{Larsson2011}. We
  note that the change in morphology and the late-time brightening is
  observed in all five broad HST filters between $2,000$ and
  $10,000\ \rm{\AA}$, with the largest increase in flux occurring at the shortest wavelengths.}

\item{The SINFONI images of the $1.644~\mu$m [Si~I]+[Fe~II]   emission
  show a more uniform morphology, with the brightest regions appearing close to the faintest, central regions of the HST images. This may be due to a remaining significant contribution from radioactive decay in the inner part of the ejecta. However, because of the high abundances of Si and Fe in the core, a similar morphology would also be expected if there is a significant hard X-ray flux reaching the innermost ejecta.  Accurate flux measurements at these wavelengths over the coming years are needed in order to determine what the dominant energy source is. }

\item{The STIS H$\alpha$ and SINFONI $1.644~\mu$m [Si~I]+[Fe~II]  
  line profiles are similar in the sense that they show predominantly blueshifted emission in
  the north and predominantly redshifted emission in the south, showing that the ejecta have a flattened distribution fairly close to the plane of the equatorial ring. There is also a north-south asymmetry: the 
  emission in the northern half of the ejecta is concentrated around
  the plane of the ring, while the emission in the southern
  part is mostly above the ring, closer to the plane of the sky. This asymmetry may partly be due to dust absorption in the far side of the ejecta, as suggested by the smooth decrease of the redshifted emission.
  The H$\alpha$ and  $1.644~\mu$m [Si~I]+[Fe~II]   profiles differ in that the H$\alpha$ emission extends to higher velocities, as expected from the differences between the images. }

\item{Three-dimensional emissivity maps of the [Si~I]+[Fe~II]  line constructed from the SINFONI observations reveal that most of the emission from these elements is concentrated to two clumps in the velocity interval $1,500 -
  3,000\ \rm{km\ s^{-1}}$. We do not have the spatial or spectral
  resolution to estimate the size of clumps from the H$\alpha$
  spectra, but in the HST images there is substructure with projected
  velocity widths of $\sim 1,000\ \rm{km\ s^{-1}}$ (the diameter of the
  'hole') to $\sim 3,500\ \rm{km\ s^{-1}}$ (the thickest part of the ring
  structure).}

\item{We have compared our results with the explosion models by
  \cite{Hammer2010}.  While there is some qualitative agreement, we also found that the observations show a higher degree of large-scale asymmetry than the simulation.  We stress,
  however, that this explosion model was not set up specifically to model \8 and that radioactive heating was not included.  
  In terms of explosion models we also note that
  the observed distribution of emission is inconsistent with a
  jet-like structure perpendicular to the equatorial ring. However,
  this does not necessarily rule out explosion models where most of
  the energy (but not ejecta) comes out along the poles.}

\end{itemize}

\acknowledgments

We thank the anonymous referee for a careful reading of the manuscript. This work was supported by the Swedish Research Council and the
Swedish National Space Board. RPK's work was supported in part by the National Science Foundation under Grant NSF PHY11-25915 to the Kavli Institute for Theoretical Physics. Support for HST GO program numbers 02563, 03853,04445, 05480, 06020, 07434, 08243, 08648, 09114, 09428,10263, 11181, 11973, 12241 was provided by NASA through grants from the Space Telescope Science Institute,
which is operated by the Association of Universities for Research in Astronomy, Inc., under NASA contract NAS5-26555. The ground-based observations were collected at the European Organisation for Astronomical Research in the Southern Hemisphere, Chile (ESO Programme 076.DÐ0558 and 086.D-0713(C))



{\it Facilities:} \facility{HST(ACS, WFPC2, WFC3, STIS)}, \facility{VLT (SINFONI, NACO)}.




\bibliographystyle{apj}
\bibliography{jlarsson_sn87a_r1}

\clearpage


\clearpage


\clearpage

\clearpage


\clearpage

\begin{deluxetable}{ccccc}
\tabletypesize{\scriptsize}
\tablecaption{HST Imaging observations \label{imtable}}
\tablewidth{0pt}
\tablehead{
\colhead{Date} & \colhead{Days since explosion\tablenotemark{1}} & \colhead{Instrument} & \colhead{Filter} & \colhead{Exposure time}\\
 &  &  &  & \colhead{(s)}
}
\startdata
1994-09-24 & 2,770 & WFPC2 & F675W & 600 \\
1998-02-06 & 4,001 & WFPC2 & F675W & 400 \\
2000-11-13 & 5,012 & WFPC2 & F255W & 5,600 \\
2000-11-13 & 5,012 & WFPC2 & F336W & 2,000 \\
2000-11-13 & 5,012 & WFPC2 & F439W & 1,200 \\
2000-11-13 & 5,012 & WFPC2 & F555W & 800 \\
2000-11-13 & 5,012 & WFPC2 & F675W & 2,400 \\
2000-11-13 & 5,012 & WFPC2 & F814W & 800 \\
2003-11-28 & 6,122 & ACS HRC & F625W & 800 \\
2006-12-06 & 7,226 & ACS HRC & F625W & 1,200 \\
2009-12-12 & 8,328 & WFC3 & F225W & 800 \\
2009-12-12 & 8,328 & WFC3 & F336W & 800 \\
2009-12-12 & 8,328 & WFC3 & F438W & 800 \\
2009-12-12 & 8,328 & WFC3 & F555W & 400 \\
2009-12-12 & 8,328 & WFC3 & F625W & 3,000 \\
2009-12-12 & 8,328 & WFC3 & F814W & 400 \\
\enddata
\tablenotetext{1}{1987 February 23}
\end{deluxetable}

\begin{deluxetable}{cccccc}
\tabletypesize{\scriptsize}
\tablecaption{HST STIS observations \label{stistable}}
\tablewidth{0pt}
\tablehead{
\colhead{Date} & \colhead{Days since explosion\tablenotemark{1}} & \colhead{Grating} & \colhead{Slit width}  & \colhead{Slit PA } & \colhead{Exposure time}\\
 &  &  & \colhead{(arcsec)}  & \colhead{(degrees east of north)}  & \colhead{(s)}
}
\startdata
1999-02-21 & 4,381 & G750L & 0.5 & 25.6 &  10,500 \\
1999-08-30 & 4,571 & G750M\tablenotemark{2} & 0.1 & 27.0 &  7,804\\
1999-08-30 & 4,571 & G750M\tablenotemark{2}  & 0.1 & 27.0 &  7,804\\
1999-08-31 & 4,572 & G750M\tablenotemark{2}  & 0.1 & 27.0 &   7,804\\
2004-07-18 & 6,355 & G750L & 0.2 & 179.7 &  5,468\\
2004-07-18 & 6,355 & G750L & 0.2 & 179.7 &  5,468\\
2004-07-23 & 6,360 & G750L & 0.2 & 179.7 &   5,468\\
2010-01-31 & 8,378 & G750L & 0.2 & 179.7  &  14,200\\
\enddata
\tablecomments{The slit
positions for day 4,571/4,572 are shown in Fig.~\ref{regions99m} and the
slit positions for days 6,355/6,360 and 8,378 are shown in
Fig.~\ref{regions0410}.}
\tablenotetext{1}{1987 February 23}
\tablenotetext{2}{Central wavelength setting 6581 \AA}
\end{deluxetable}

\end{document}